\documentclass[aps,showpacs,preprintnumbers,amsmath,nofootinbib,amssymb]{revtex4}
\usepackage[dvips]{epsfig}

\newcommand{\nc}[1]{\newcommand{#1}}

\nc{\its}[1]{\itshape #1 \upshape}
\nc{\mc}[3]{\multicolumn{#1}{#2}{#3}}

\nc{\bc}{\begin{center}}
\nc{\ec}{\end{center}}
\nc{\ig}[1]{\bc \includegraphics{#1} \ec}

\nc{\bo}[1]{\mbox{\boldmath \( #1 \! \! \)  \unboldmath}}
\nc{\be}{\begin{eqnarray}}
\nc{\ee}{\end{eqnarray}}
\nc{\bew}{\begin{eqnarray*}}
\nc{\eew}{\end{eqnarray*}}
\nc{\bs}{\begin{subeqnarray}}   
\nc{\es}{\end{subeqnarray}}     
\nc{\nnn}{\nonumber \\}
\nc{\f}[2]{\frac{#1}{#2}}
\nc{\td}[2]{\f{d #1}{d #2}}
\nc{\pd}[2]{\f{\partial #1}{\partial #2}}
\nc{\suli}{\sum\limits}
\nc{\proli}{\prod\limits}
\nc{\ili}{\int\limits}
\nc{\sr}[2]{\stackrel{#1}{#2}}
\nc{\dps}{\displaystyle}
\nc{\ket}[1]{\left| #1 \right>}
\nc{\bra}[1]{\left< #1 \right|}
\nc{\bracket}[2]{\left< #1 \right| \left. \! #2 \right>}
\nc{\norm}[1]{\left\| #1 \right\|}
\nc{\lndm}[1]{\pd{^{#1} \ln{\det{M}}}{\mu^{#1}}}
\nc{\pdmm}[1]{M^{-1} \pd{^{#1} M}{\mu^{#1}}}
\nc{\pdm}{M^{-1}\pd{M}{\mu}}
\nc{\trac}[1]{\mbox{Tr}\left(#1\right)}
\nc{\hm}{\hat{m}}

\def\lsim{\raise0.3ex\hbox{$<$\kern-0.75em\raise-1.1ex\hbox{$\sim$}}}
\def\gsim{\raise0.3ex\hbox{$>$\kern-0.75em\raise-1.1ex\hbox{$\sim$}}}

\begin{document}

\title{Study of the finite temperature transition in 3-flavor QCD using the R and RHMC algorithms}

\author{M. Cheng$^{\rm a}$, N. H. Christ $^{\rm a}$, M.A. Clark$^{\rm b}$, 
J. van der Heide$^{\rm d}$,
C. Jung$^{\rm c}$, F. Karsch$^{\rm c,d}$, O. Kaczmarek$^{\rm d}$,\\ 
E. Laermann$^{\rm d}$, R. D. Mawhinney$^{\rm a}$, C. Miao$^{\rm d}$,
P. Petreczky$^{\rm c,e}$, K. Petrov$^{\rm f}$,
C. Schmidt$^{\rm c}$, W. Soeldner$^{\rm c,d}$, and T. Umeda$^{\rm c}$
}

\affiliation{
$^{\rm a}$ Physics Department, Columbia University, New York, NY 10027, USA\\
$^{\rm b}$ Center for Computational Science, Boston University, Boston, MA 02215\\
$^{\rm c}$Physics Department, Brookhaven National Laboratory, 
Upton, NY 11973, USA \\
$^{\rm d}$Fakult\"at f\"ur Physik, Universit\"at Bielefeld, D-33615 Bielefeld,
Germany\\
$^{\rm e}$ RIKEN-BNL Research Center, Brookhaven National Laboratory, 
Upton, NY 11973, USA \\
$^{\rm f}$Niels Bohr Institute, University of Copenhagen, 
Blegdamsvej 17, DK-2100 Copenhagen, Denmark\\
}

\date{\today}
\preprint{BNL-NT-06/45}
\preprint{BI-TP 2006/40}
\preprint{CU-TP-1163}

\begin{abstract}
We study the finite temperature transition in  
QCD with three flavors of equal masses using the R and RHMC algorithm on lattices
with temporal extent $N_{\tau}=4$ and $6$.
For the transition temperature in the continuum limit we find $r_0 T_c=0.429(8)$ for the 
light pseudo-scalar mass
corresponding to the end point of the 1st order transition region.
When comparing the results obtained with the R and RHMC algorithms for
p4fat3 action we see no significant step-size errors down to a lightest
pseudo-scalar mass of $m_{ps} r_0=0.4$. 
\end{abstract}


\maketitle

\section{Introduction}
\label{intro}
Lattice QCD has established the existence of a transition from hadron gas to a 
new state of strongly interacting matter where quarks and gluons are no longer confined
inside hadrons and which is usually called the quark gluon plasma \cite{cargese,reviews}. 
The nature of this 
transition depends on the quark content and quark masses. For infinite or very large quark
masses the transition is a 1st order deconfining transition. In the opposite case of zero
quark masses one may have a 2nd order chiral phase transition for 2 flavors or
a 1st order chiral phase transition for 3 flavors. For intermediate masses the transition is just
a rapid crossover, meaning that thermodynamic quantities change very rapidly in a narrow
temperature interval. The boundary of the 1st order transition region of 3 flavor QCD as a function of mass 
has been studied using improved (p4) \cite{karschlat03} and standard staggered actions
\cite{christian,norman}. There is a significant discrepancy regarding the value of the quark masses, or
equivalently the value of the pseudo-scalar meson masses where the transition changes
from crossover to 1st order. With an improved action it was found that the 1st order transition ends
for a pseudo-scalar meson mass of about $70$ MeV, while with the standard action it ends for
pseudo-scalar meson masses of about $190$ MeV \cite{christian,phil2} or larger \cite{norman}.

It has been observed that the pressure and energy density normalized by its ideal gas
value shows almost the same behavior as a function of $T/T_c$ for SU(3) pure gauge theory,
2 flavor, 2+1 flavor and 3 flavor QCD \cite{cargese}. Thus  flavor and quark mass dependence
of these quantities in the first approximation, is determined by flavor and quark mass dependence of 
transition temperature $T_c$. Therefore it is very instructive to
study the flavor dependence of the transition temperature. Such a study has been 
performed in Ref. \cite{karsch01} on lattices with temporal extent $N_{\tau}=4$ using the
improved staggered p4 action.

In the past most simulations with 2 and 3 flavors of staggered fermions have been done using the
hybrid molecular dynamics R (HMDR) algorithm \cite{hybrid}, often called simply the R-algorithm. It has finite step-size errors
of ${\cal O}(dt^2)$ where a step-size $dt$ is used in  the molecular dynamics evolution. Recently the rational hybrid Monte Carlo (RHMC) algorithm has been
invented which allows simulations of  theories with fractional powers of the fermion determinant, for example
2 and 3 flavors of staggered fermions, without finite step-size errors \cite{rhmc}. Therefore the most recent
thermodynamics studies use the RHMC algorithm \cite{aoki06,philipsen06,SK06,us}.
It has been observed in Ref. \cite{philipsen06} that the use of the exact RHMC algorithm
reduces the value of the critical quark mass where the transition turns to 1st order by $25 \%$ in the
case of the standard staggered action.

The purpose of this paper is twofold. First we would like to study the transition in 3 flavor QCD,
extending the previous studies to smaller quark masses and smaller lattice spacings using the 
improved p4 staggered fermion action. Second we would like to compare the R-algorithm with the
new RHMC algorithm. 
The rest of the paper is organized as follows. In section II
we discuss the calculational set-up. In section III we show our results for finite temperature calculations.
Section IV discusses the zero temperature simulations needed to set the scale and the transition
temperature in physical units. In section V we present a comparison of the R and RHMC algorithms.
Finally, section VI contains our conclusions. A technical discussion of different fat link actions is
given in  Appendix A. In Appendix B we discuss the eigenvalues and eigenvectors of the p4 action in the free field limit.

\section{Lattice Formulation and setup}
\label{setup}

Most of the simulations discussed in this paper were done using the 
p4fat3 action, which is also simply called the p4 action \cite{karsch01}.
To improve the rotational symmetry, which is violated on the lattice, bent 3-link terms are
added to the 1-link term of the standard staggered action \cite{Heller}. 
Properly chosen coefficients of the 1-link and the 3-link terms  can eliminate,
at tree level, the  ${\cal O}(a^2)$ errors in the dispersion relation for staggered fermions \cite{Heller}.
The violation of flavor symmetry which is present in the staggered fermion formulation 
can be significantly reduced by replacing the normal link in the 1-link term by a fat link
which is a sum a of normal link and 3-link staples \cite{tb97}. This 3-link fattening is the origin of the name p4fat3.
Though this type of fat link action is a big improvement over the standard staggered
fermion action, further improvement of the flavor symmetry can be obtained by adding 5-
and 7-link staples \cite{kostas}. In fact one can eliminate the effect of flavor symmetry breaking at
order ${\cal O}(g^2 a^2)$ using a suitable combination of 3-, 5- and 7- link staples leading to
what is called the fat7 action \cite{kostas}.  We have also done calculations with the p4
action with fat7 fat links. Unfortunately it turns out that on the  coarse lattices used in our study of  3 flavor QCD
thermodynamics this action has some undesirable features. It leads to the occurrence of a bulk transition, which
we will discuss in more detail in Appendix A.

To study staggered fermions with less than four  flavors,  we use the rooting
procedure, i.e. each fermion flavor is represented by  $({\rm det} \mathcal{M})^{1/4}$, where $\mathcal{M}$ is the staggered fermion Dirac operator. 
For a recent discussion of this procedure see Ref. \cite{sharpe}.
Most of our simulations have been done using the standard R-algorithm \cite{hybrid}. As in Ref. \cite{karsch01}
the step-size of the molecular dynamics evolution was set to $dt=m/2.5$ for the staggered quark mass $m$.

We also performed calculations with the RHMC algorithm. In this algorithm an optimal rational
approximation is used to evaluate the fractional power of the determinant \cite{rhmc}.  
More precisely one finds the optimal approximation for $(\mathcal{M}^{\dagger} \mathcal{M})^{\nu}$ where $\mathcal{M}=2 m +D$ is
the usual staggered fermion matrix and for the three flavors $\nu=3/8$. Using 
sufficiently high order polynomials   the rational approximation can be made arbitrarily precise for
the given spectral range of the fermion operator. 
For the standard staggered fermion action the spectral range of $\mathcal{M}^{\dagger} \mathcal{M}$ is well 
known, the smallest eigenvalue of this matrix is $4 m^2$. The largest eigenvalue can be estimated in the free field limit to
be  $\lambda^2_{max}=16+4 m^2$.
For the case of the p4 staggered action the smallest eigenvalue is the same, while the
largest eigenvalue in the free case is $\lambda_{max}^2=50/9+4 m^2$. In appendix B we give
the derivation of this result.
It turns out that
in all our simulations the largest eigenvalue was smaller than 5.0, so we choose 
$\lambda_{max}^2=5.0$
as the upper limit on it. 
For the range of the quark masses studied by us, which include quark masses
as light as 1/20th of the strange quark mass, it is sufficient to use polynomials 
of degree 12 to achieve
machine precision with the rational approximation. 
Therefore in the Metropolis accept/reject step we used
polynomials of degree 12. In the molecular dynamics evolution we used a less 
stringent approximation of
the determinant since any errors in the evolution, including the $dt^2$ 
step-size errors, are eliminated by the
accept/reject step. It has been found that in most cases it is sufficient to use 
polynomials of degree 5-6
without compromising the acceptance rate. 
Furthermore, the stopping criteria for the conjugate gradient inversions can be relaxed
to $10^{-5}$ in the molecular dynamics evolution without significant 
effect on the acceptance rate. In the Monte-Carlo accept/reject step we typically use  $10^{-12}$ for
the conjugate gradient stopping criteria. The length of the trajectory was $\tau_{MD}=0.5$
in units of molecular dynamics time.

The gauge fields and quarks contribute different amounts to the force in 
the molecular dynamics evolution.
The contribution of the gauge fields is larger than that of quarks. On the other hand   
the cost of the evaluation of the force coming from the gauge fields is small. 
The opposite is true
for the part of the force coming from the fermions. Therefore it is reasonable to 
integrate the gauge force on finer molecular dynamics time scale than the fermion force \cite{sextwein}.
In our simulations we typically used 10 gauge field updates per fermion update. The 3 flavors
of staggered fermions are simulated as 1+1+1, i.e. we used a factor $({\rm det} \mathcal{M}^{\dagger} \mathcal{M})^{1/8}$  
for each fermion flavor. Although this increases the number of inversions, the reduced force allows
for a larger step-size $d t$ for the same acceptance rate. The step-size $dt$ was chosen such that 
the acceptance is about $70 \%$. To achieve this, the stepsize was typically of the order
of the strange quark mass used in our 2+1 flavor study \cite{us}.

\begin{table}[t]
\begin{center}
\vspace{0.3cm}
\begin{tabular}{|c|r|c|c|c|}
\hline
$N_\tau$ &   $m$     & $N_\sigma~~$     &\# $\beta$ values & max. no. of traj. \\
\hline
4             & 0.100      & 16                     & 4                         &  42000 \\
~             & 0.050      & 8,~16                & ~8,~14                &  4130,~ 2650\\
~             & 0.025      & 8,~16                & ~8,~~9                &  6250, ~8650 \\
~             & 0.010      & 8,~16                & ~9,~~6                &  2460, ~4660 \\
~             & 0.005      & 12,~16              & ~11,~8                &  1360,~3000  \\
\hline
6            & 0.100       & 16                       &  16                    & 6000             \\
~            & 0.050       & 16                       &  14                    & 7900             \\
~            & 0.020       & 16                       &  10                    & 16000           \\
~            & 0.010       & 16                       &   9                     & 10900           \\
\hline
\end{tabular}
\end{center}
\caption{
Parameters of the numerical simulations
}
\label{tab:parameter}
\end{table}

\section{Finite temperature simulations}

Our studies of the QCD transition at finite temperature have been
performed on lattices of size $N_\sigma^3\times N_\tau$. The 
lattice spacing, $a$, relates the spatial ($N_\sigma$) and temporal 
($N_\tau$) size of the lattice to the physical volume 
$V= (N_\sigma a)^3$ and temperature $T=1/N_\tau a$, respectively.
The lattice spacing, and thus the temperature, is controlled by the
gauge coupling, $\beta = 6/g^2$, as well as the bare quark masses.
The parameters of our finite temperature simulations are given in Table \ref{tab:parameter}.
We extended the results of Ref. \cite{karsch01} in two respects.
Compared to Ref. \cite{karsch01} we have added a smaller mass value $m=0.005$ for $N_{\tau}=4$ and
extended the runs at larger quark masses to achieve a better statistical accuracy.
In addition we have studied the finite temperature transition on $N_{\tau}=6$ 
lattices for two values of the quark mass $m$. All the results presented in this section have been
obtained with the R algorithm.

As mentioned in the introduction, in 3 flavor QCD for small quark mass we have a 1st order phase transition
which turns into rapid crossover at the quark mass corresponding to the light pseudo-scalar mass
of about $70$ MeV \cite{christian}.
The transition is signaled by a 
rapid change in bulk thermodynamic observables (energy density, pressure)
as well as in the chiral condensates and the Polyakov loop expectation value,
\begin{eqnarray}
\frac{\langle \bar{\psi}\psi \rangle}{T^3} &=& \frac{1}{3} \frac{1}{VT^2}
\frac{\partial \ln Z}{\partial m}
= \frac{N_\tau^2}{4 N_\sigma^{3}} 
\left\langle {\rm Tr}\; M^{-1}(m)\right\rangle
\label{orderparameter1}  
, \\
\langle L\rangle &=& \left\langle \frac{1}{3N_\sigma^{3}} 
{\rm Tr}\sum_{\bf x} 
\prod_{x_0 = 1}^{N_\tau} U_{(x_0, {\bf x}),\hat{0}} \right\rangle \; , 
\label{orderparameter}
\end{eqnarray}
which are order parameters for a true phase transition in the zero 
and infinite quark mass limit, respectively. 
Note that we have defined the chiral condensate per flavor
degree of freedom, hence the factor 1/3 in Eq.  (\ref{orderparameter1}). 

\begin{figure}
\includegraphics[width=7cm]{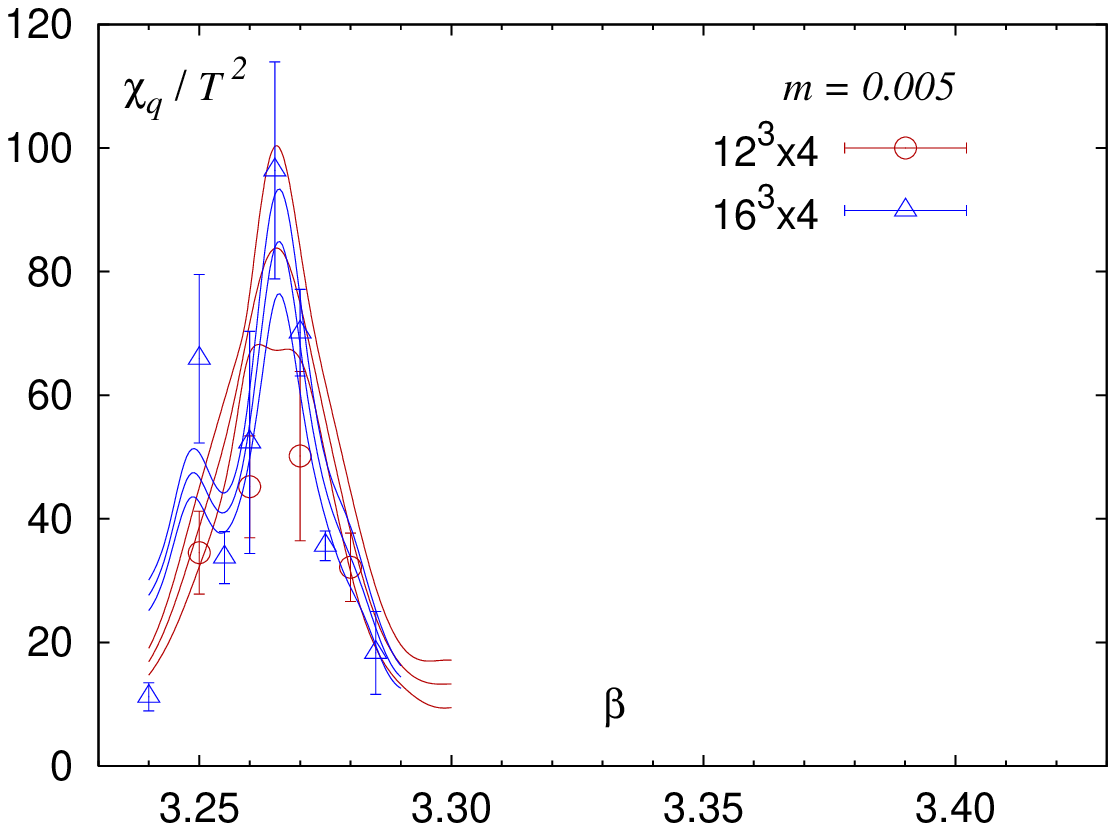}
\includegraphics[width=7cm]{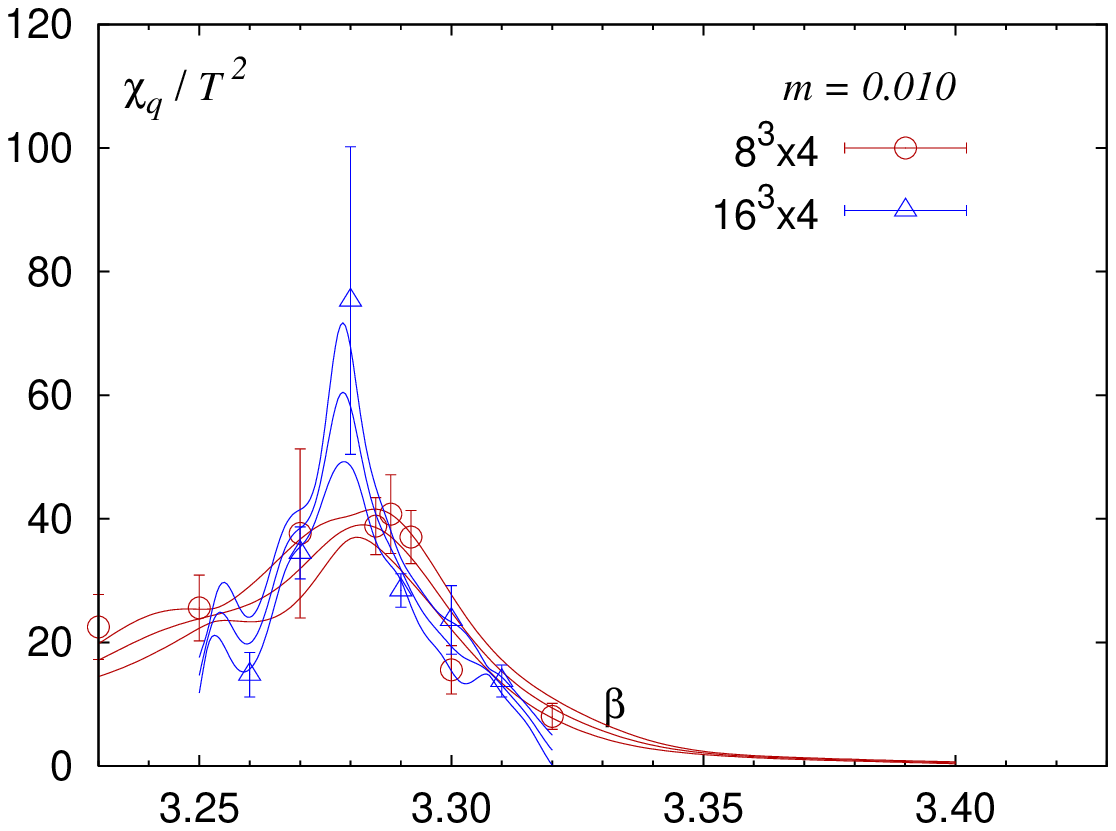}
\includegraphics[width=7cm]{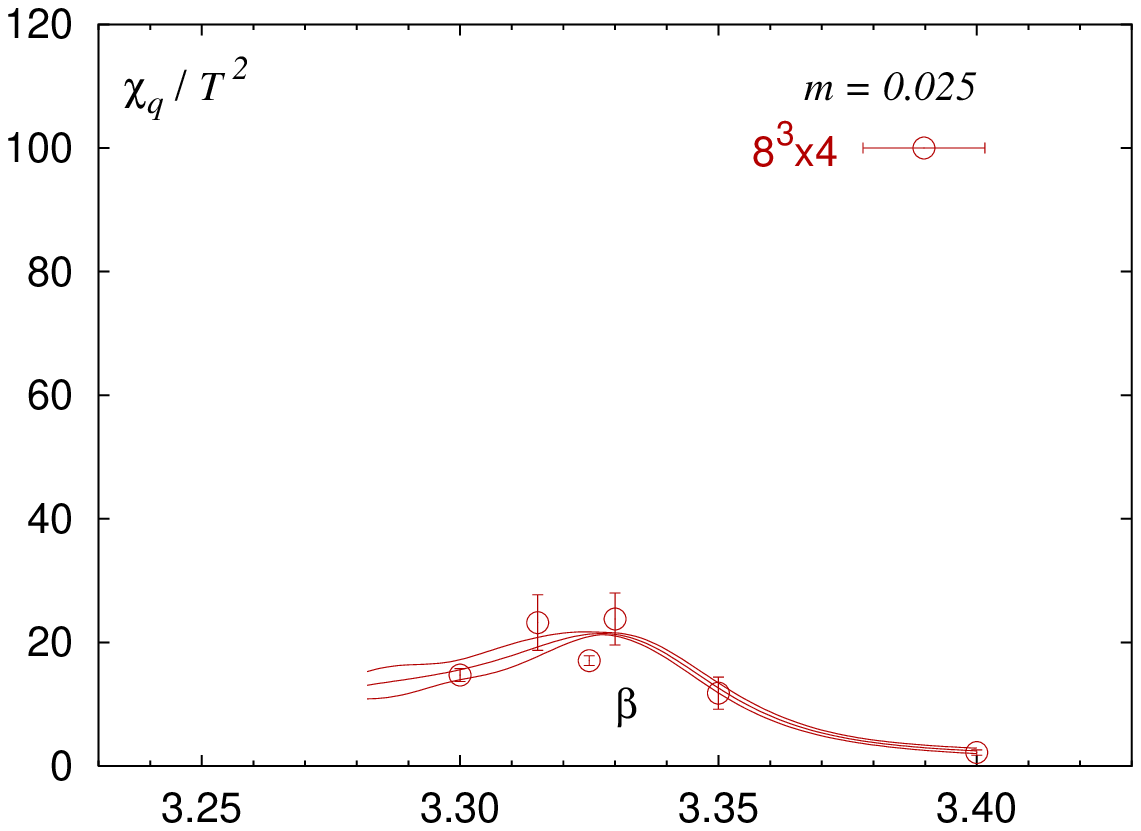}
\includegraphics[width=7cm]{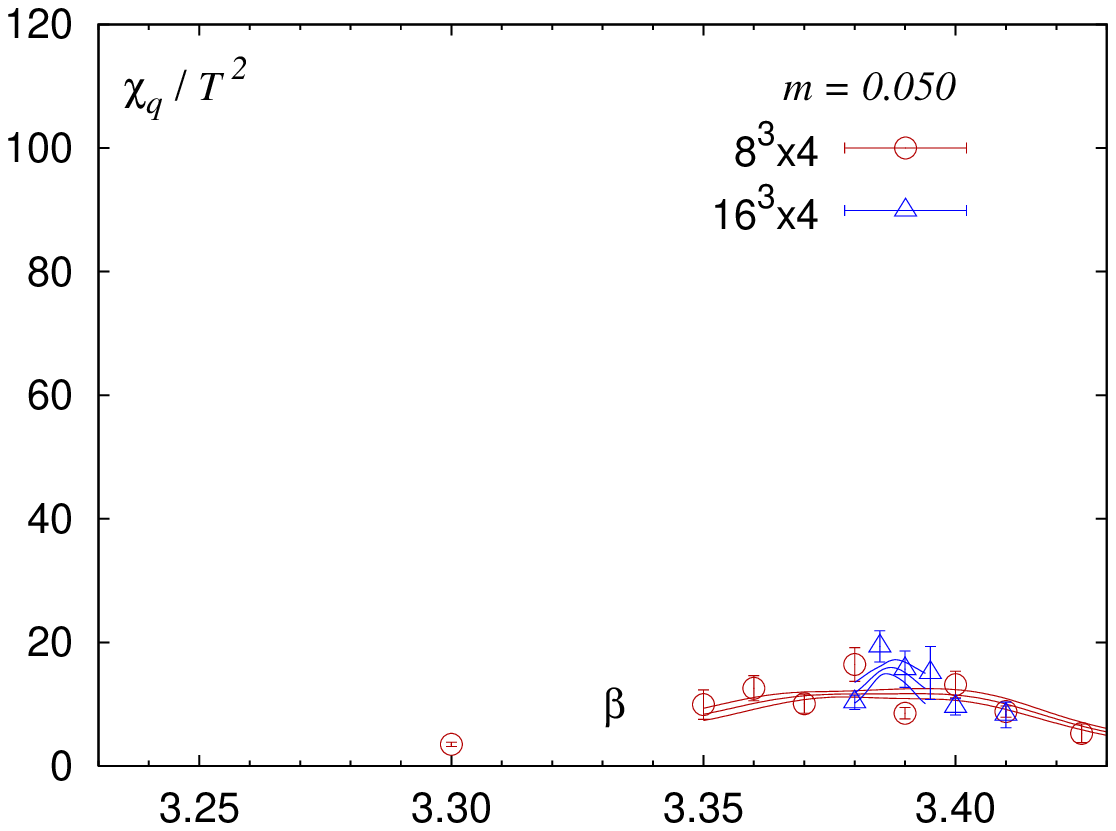}
\caption{The disconnected part of the chiral susceptibility calculated 
on $N_{\tau}=4$ lattices at different quark masses.}
\label{fig:suscnt4}
\end{figure}

\begin{figure}
\includegraphics[width=8cm]{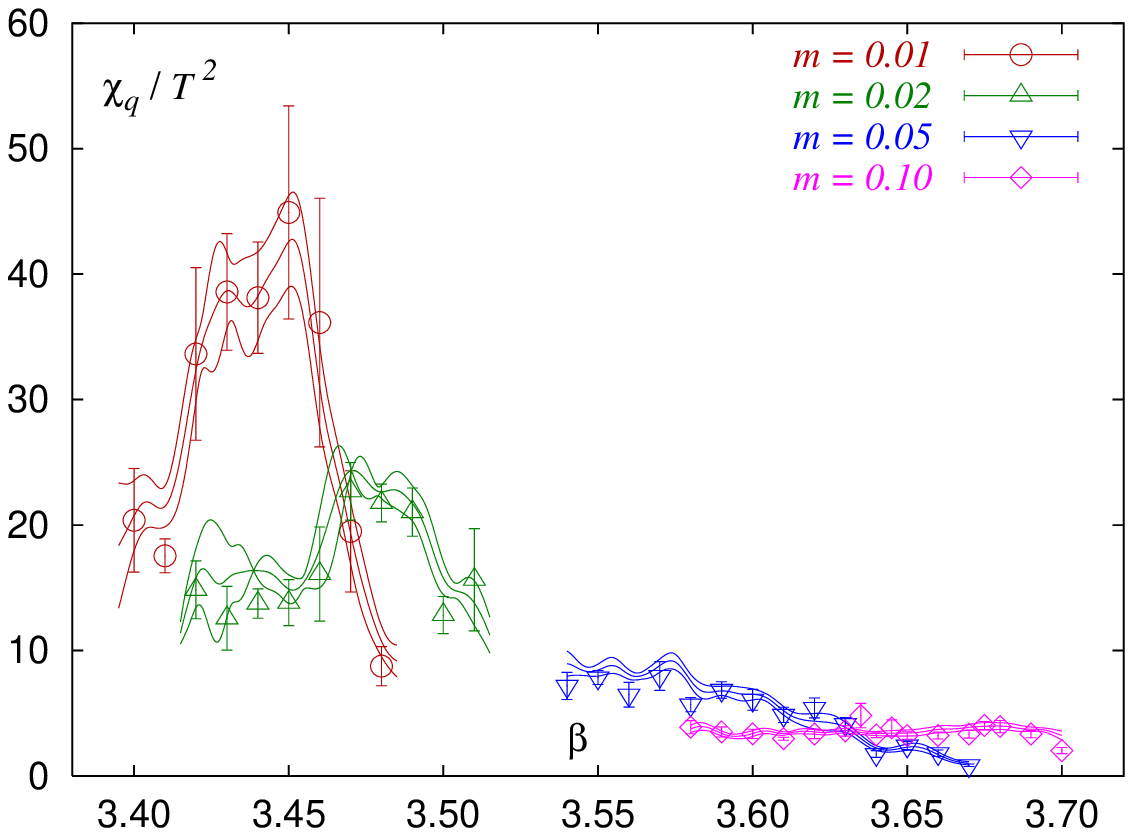}
\includegraphics[width=8cm]{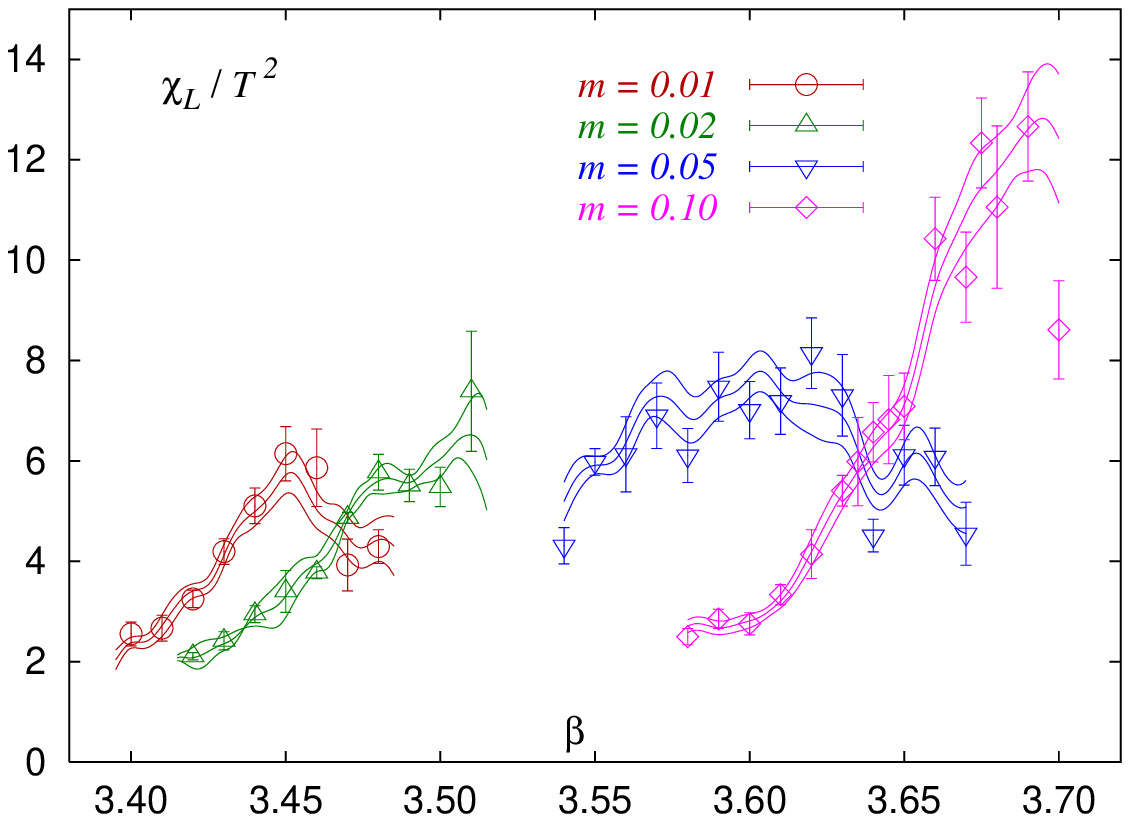}
\caption{The disconnected part of the chiral susceptibility (left) and the
Polyakov loop susceptibility (right) calculated on $16^3 \times 6$ lattices.}
\label{fig:suscnt6}
\end{figure}

We use  the Polyakov loop susceptibility as well as the disconnected part 
of the chiral susceptibility to locate the transition temperature.
\begin{eqnarray}
\chi_L &\equiv& N_\sigma^{3}  \left( 
\langle  L^2 \rangle - \langle L \rangle^2 \right) \; ,
\label{sus_L}\\
\frac{\chi_q}{T^2} &\equiv& \frac{N_\tau}{16N_\sigma^{3}} \left( 
\left\langle \left( {\rm Tr}\; M^{-1}(m)\right)^2 \right\rangle -
\left\langle {\rm Tr\;} M^{-1}(m)\right\rangle^2\right) 
. \label{sus_chi}
\label{sus_m}
\end{eqnarray}
We calculate the value of the Polyakov loop at the end of every trajectory.
For each tenth trajectory we calculate the value of $\bar \psi \psi$ and $\chi_q$ using ten Gaussian
random vectors.
In Fig. \ref{fig:suscnt4} we show the disconnected part of the chiral susceptibility
calculated on our $N_{\tau}=4$ lattice with different spatial volumes at different
quark masses.
In Fig. \ref{fig:suscnt6} we show the Polyakov loop $\chi_L$ and chiral $\chi_q$ susceptibilities
calculated on $16^3 \times 6$ lattice. 
The location of peaks in the susceptibilities
has been determined using Ferrenberg-Swedsen re-weighting for several values $\beta$
in the vicinity of the transition.  Errors on the peak location have been obtained from
a jackknife analysis where Ferrenberg-Swedsen re-weighting has been performed on different
sub-samples. The resulting pseudo-critical couplings are shown in Table \ref{tab:betac}.
In finite volume the pseudo-critical couplings $\beta_c$ determined from the Polyakov loop
correlator and chiral susceptibility are generally different. In the case of the crossover
this difference can persist even in the infinite volume limit. From  Table \ref{tab:betac} we see
that in most cases the two pseudo-critical couplings are identical within statistical errors even for
small volume $8^3 \times 4$. 
The cases where this difference is the largest are the cases where $\beta_{c,q}$ has large
statistical errors. For example for the $16^3\times 6$ lattice and $m=0.05$ we find
$\beta_{c,L}-\beta_{c,q}=.0113(328)$.
Therefore we have also calculated the weighted average of  $\beta_{c,L}$ and $\beta_{c,q}$
which is shown in the last column of Table \ref{tab:betac} together with the corresponding
error. This error is calculated from the statistical errors and the difference between the central
values added quadratically. 
The difference in pseudo-critical couplings determined on $8^3 \times 4$ and
$16^3 \times 4$ lattices is typically small, indicating small finite volume effects.
As a general tendency the pseudo-critical coupling $\beta_c$ shifts toward smaller values
with increasing volume.  
In agreement with earlier calculations we find that the position of
peaks in $\chi_q$ and $\chi_L$ show only little volume dependence and that
the peak height changes only little, although the maxima become somewhat more
pronounced on the larger lattices. This is consistent with the transition being
a crossover rather than a true phase transition in the infinite volume limit
for the range of quark masses explored by us.
\begin{table}
\begin{tabular}{|c|c|c|c|c|c|}
\hline
$N_{\tau}$   &   $m$     & $N_{\sigma}$  &   $\beta_{c,L}$ [from $\chi_L$]   &    $\beta_{c,q}$ [from $\chi_q$]   & $\beta_c$ [averaged] \\
\hline
  4                &   0.100   & 16                    &   3.4800(27)                               &    3.4804(24)                                &  3.4802(18)   \\
                    &   0.050   & 16                    &   3.3884(32)                               &    3.3862(47)                                &  3.3877(34)   \\
                    &               & 8                      &   3.4018(35)                               &    3.3930(201)                              &  3.4015(94) \\
                    &   0.025   & 8                      &   3.3294(27)                               &    3.3270(28)                                &   3.3283(31)   \\
                    &   0.010   & 16                    &   3.2781(7)                                 &    3.2781(4)                                  &   3.2781(3)     \\
                    &               & 8                      &   3.2858(71)                               &    3.2820(61)                                &   3.2836(60) \\
                    &   0.005   & 16                    &   3.2656(13)                               &    3.2678(12)                                &   3.2667(24)  \\
                    &               & 12                    &   3.2659(13)                               &    3.2653(12)                                &   3.2656(10)  \\
\hline
6                  &   0.200   & 16                     &  3.8495(11)                               &    3.9015(279)                              &    3.8495(520) \\
                    &   0.100   & 16                     &  3.6632(55)                               &    3.6855(105)                              &    3.6680(228) \\
                    &   0.050   & 16                     &  3.6076(24)                               &    3.6189(328)                              &    3.6077(115) \\
                    &   0.020   & 16                     &  3.4800(110)                             &    3.4800(80)                                &    3.4800(65)  \\
                    &   0.010   & 16                     &  3.4518(50)                               &    3.4510(83)                                &    3.4516(44) \\
\hline
\end{tabular}
\caption{Critical couplings determined from the location of peaks in the Polyakov loop susceptibility as well
as in the disconnected parts of the chiral susceptibilities. The last column gives the average of $\beta_{c,L}$
and $\beta_{c,q}$ with combined statistical and systematic errors.}
\label{tab:betac}
\end{table}

\section{Zero temperature calculations and the transition temperature}

In order to determine the transition temperature in units of some physical 
quantity we performed
zero temperature calculations on $16^3 \times 32$ lattices in the vicinity of 
the pseudo-critical 
coupling $\beta_c$. 
The parameters of these calculations together with the accumulated statistics 
are summarized in Table \ref{tab:T=0}.
We have calculated the static quark potential
and meson correlators on each 10th trajectory generated.

The static potential has been calculated using  the ratios of the 
Wilson loops at two neighboring time-slices and extrapolating them to infinite time
separation with the help of constant plus exponential form.
The spatial transporters in the Wilson loop have been constructed 
from spatially
smeared links with APE smearing. The weight of the 3 link staple was $\gamma=0.4$
and we used ten steps of APE smearing. From the 
static potential we have determined the string tension and the Sommer parameter $r_0$ defined
as \cite{Sommer}  
\begin{equation}
r^2\frac{{\rm d} V_{\bar{q}q}(r)}{{\rm d}r}\biggl|_{r=r_0} = 1.65.
\label{r0}
\end{equation}
When extracting $r_0$ and the string tension  on coarse lattices, such as the ones 
used in thermodynamics studies, the
violation of rotational symmetry has to be taken into account. We do this using the procedure described
in detail in our recent paper \cite{us}.
The value of Sommer scale and the string tension are given in Table \ref{tab:T=0} for different quark masses.
Having determined $r_0$ for different gauge couplings and quark masses allows us to perform interpolations
of $r_0/a$ in these parameters. As in Ref. \cite{us} we use the following renormalization group inspired
interpolation ansatz \cite{allton}
\begin{equation}
(r_0/a)^{-1} =
R(\beta) (1 +B \hat{a}^2(\beta) + C \hat{a}^4(\beta))
{\rm e}^{A (2 m_l + m_s)+D}
\; .
\label{interpolate}
\end{equation}
Here $R(\beta)$ is 2-loop beta function of 3 flavor QCD.
In the interpolation we also used the values of $r_0/a$ determined in our 2+1 flavor study \cite{us}
in addition to those shown in Table \ref{tab:T=0}, giving $A=1.45(5)$, $B=1.20(17)$,  $C=0.21(6)$
and $D=2.41(5)$ with $\chi^2/dof=0.9$. This was the reason for  using the notation
$m_l$ and $m_s$  for the light and the strange quark mass in Eq. (\ref{interpolate}) . For 3 degenerate flavors of course $m_s=m_l=m$.
In  Table \ref{tab:T=0}  we also show the values of $r_0/a$ obtained from this ansatz for each of
the parameter sets.

Meson masses have been calculated using the four local staggered meson operators. We used 
point-wall meson correlation functions with a $Z_2$ wall source. To extract the meson masses
from
the correlation functions we used a double exponential ansatz which 
takes into account the two lowest
states with opposite parity.  The two lowest pseudo-scalar meson masses 
as well as the lightest vector meson mass
are shown in  Table \ref{tab:T=0}. The breaking of the flavor symmetry can be 
quantified by the quadratic
splitting of the pseudo-scalar masses: 
$\Delta_{ps}=(m_{ps2}^2-m_{ps}^2) r_0^2$, where $m_{ps2}$ is the mass of the 
lightest pseudo-scalar non-Goldstone meson that is present with staggered fermions. 
This quantity should
be quark mass independent for sufficiently small  quark masses and should vanish 
as ${\cal O}(a^2)$ when the 
continuum limit ( $a \rightarrow 0$) is approached. 
In the last column of Table \ref{tab:T=0} we show the value 
of  $\Delta_{ps}$ from our scale setting run for $N_{\tau}=4$ and $N_{\tau}=6$.  
As we see from the table this quantity does
not decrease quite as fast as $a^2$. This is an indication that on the coarse 
lattice, corrections to asymptotic scaling are
still important.

\begin{table}
\begin{tabular}{|c|c|c|c|c|c|c|c|c|c|}
\hline
$\beta$ &  $m$   & \# traj    &  $m_{ps}$  & $m_{ps2}$ & $m_V$     & $r_0$      & $(r_0)_{smooth}$ &  $\sqrt{\sigma}$  &  $\Delta_{ps}$ \\
\hline     
3.3877  &  0.050 &  7800    &  0.7084(1)  & 1.094(7)     & 1.310(20)   & 2.066(7)[7]       &  2.061            &   0.552(12)[12]         &   2.97(7)   \\
3.3270  &  0.025 &  12000  &  0.5118(3)  & 0.998(24)   & 1.222(32)   & 1.982(14)[13]   &  1.989            &   0.564(11)[11]         &   2.90(20) \\
3.2680  &  0.005 &  1500    &  0.2341(9)  & 0.860(90)   & 1.250(50)   & 1.888(15)[9]     &  1.888            &   0.587(17)[17]         &   2.44(55)  \\
\hline
3.46345&   0.020 &  4420   &  0.4413(8)  &   0.665(5)   &  0.908(11)  &  2.797(20)[20]  &  2.813            &   0.404(6)[6]             &   1.94(9)  \\
3.4400  &   0.010 &  4290   &  0.3210(7)  &   0.594(7)   &  0.882(20)  &  2.770(13)[13]  &  2.779            &   0.405(6)[6]             &   1.92(7)  \\
\hline
\end{tabular}
\caption{Parameters of the zero temperature simulations, meson masses, the Sommer scale $r_0$ and the string tension.
Also shown is the value of $r_0$ obtained from the interpolation formula (\ref{interpolate}). The upper part of the table refers to
scale setting runs for our $N_{\tau}=4$ lattices while the lower part 
to our $N_{\tau}=6$ calculations. In the last column the splitting between the lightest non-Goldstone and the Goldstone pseudo-scalar meson masses squared is shown.
All dimensionfull quantities are given in units of the lattice spacings.
}
\label{tab:T=0}
\end{table}

\begin{figure}
\includegraphics[width=7cm]{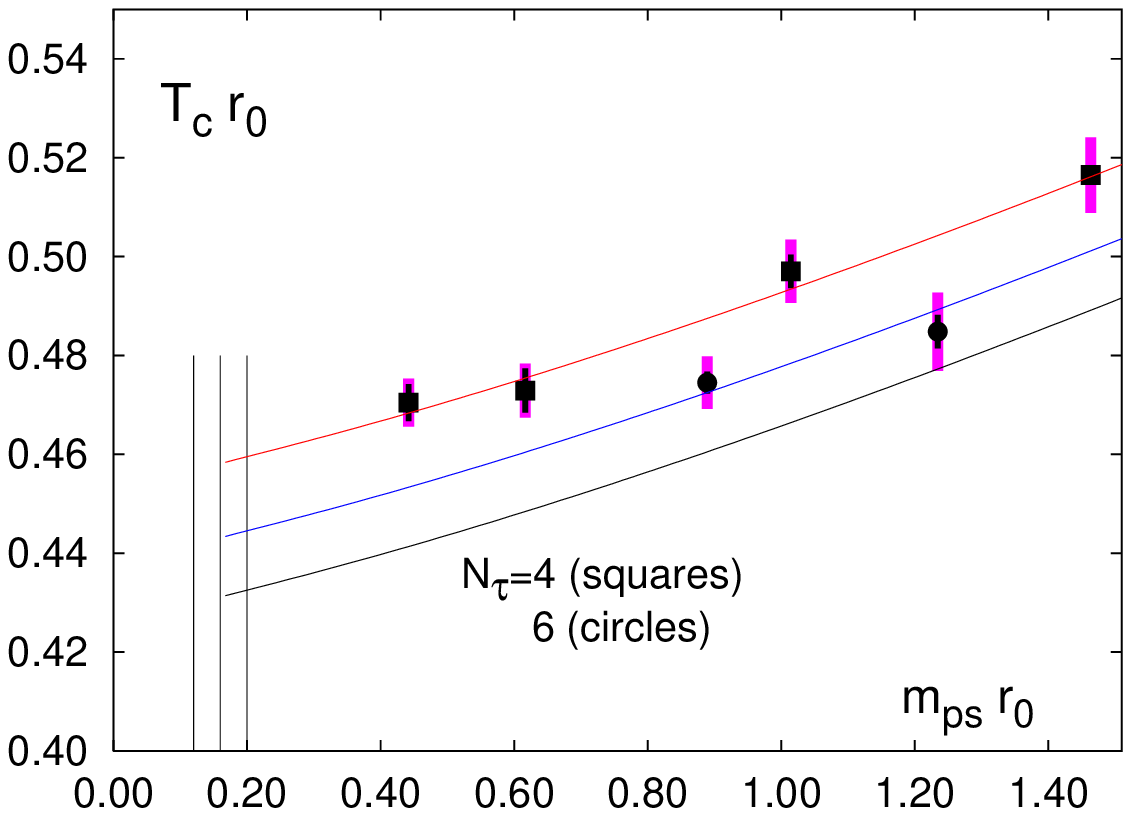}
\includegraphics[width=7cm]{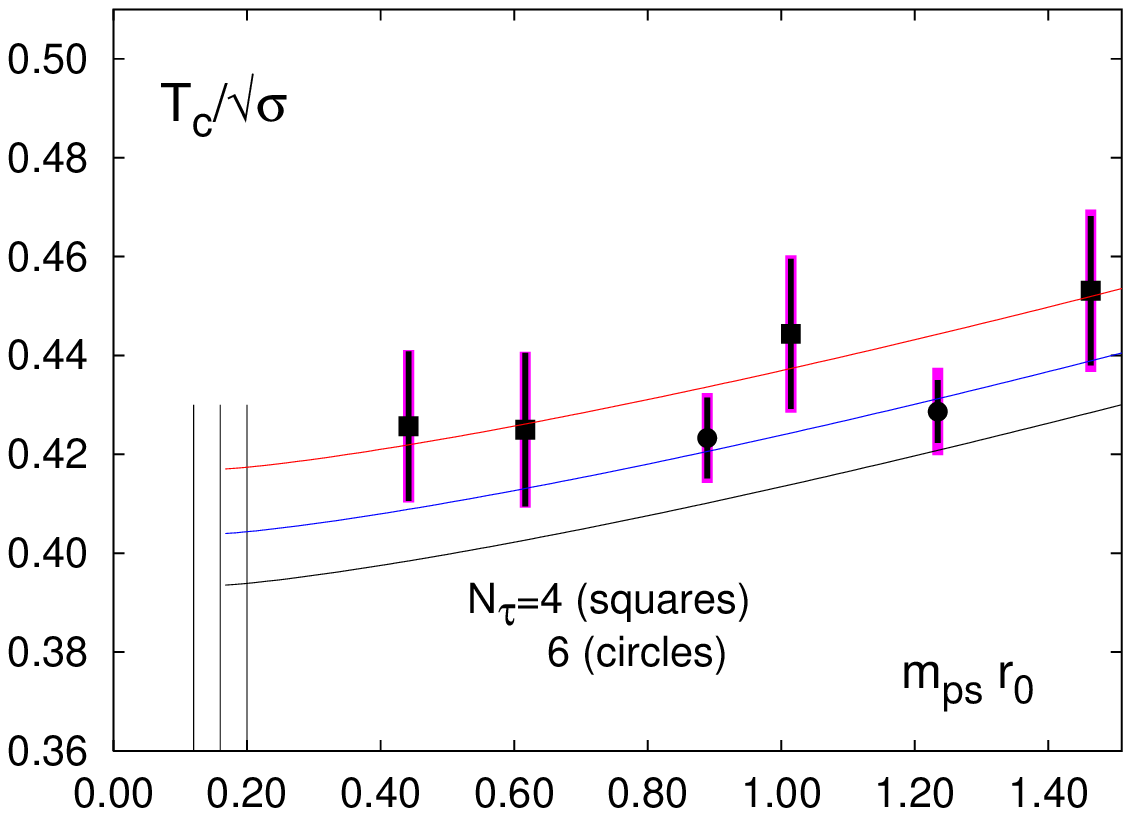}
\caption{The value of the transition temperature in units of $r_0$ (left) and in units of $\sqrt{\sigma}$ (right) calculated on 
$N_{\tau}=4$ and $N_{\tau}=6$ lattices as function of $m_{ps}$ together with continuum extrapolated value.
The vertical line and band indicates the value of $m_{ps}$ where the transition becomes 1st order.}
\label{fig:tc}
\end{figure}

With the help of the interpolation formula we can calculate the lattice spacing in units of $r_0$ and thus 
$r_0 T_c$ for different pseudo-scalar meson masses, which is shown in Fig. \ref{fig:tc}. The error in $a(\beta_c)$ results in an error in the value of
$r_0 T_c$ which is shown in Fig.   \ref{fig:tc} as thin error-bars. 
The uncertainty in $\beta_c$ itself also contribute to the uncertainty in $r_0 T_c$, which is shown
as a thick error-bar in the figure. 

If there is a critical point in the $(m,T)$-plane, 
then universality dictates that
$T_c(m)-T_c(m^e) \sim (m-m^e)^{1/(\delta \beta)}$ with $\beta$ and $\delta$ 
being critical exponents. 
In the case of three degenerate flavors the line of the 1st order transition in the 
$(m,T)$ plane ends in a critical end-point $m_e, T_c(m^e)$ 
belonging to the Z(2) universality class. For this universality class we have $\delta \beta=1.5654$. 
Therefore
we attempted a combined continuum and chiral extrapolation using the following extrapolation
ansatz 
\begin{equation}
r_0 T_c(m_{ps},N_{\tau})=r_0 T_c|_{cont}(m_{ps}^e)+ A \left ( 
\left (r_0 m_{ps} \right )^2 - \left( r_0 m_{ps,c} \right )^2 \right)^{1/(\delta \beta)} 
+ B/N_{\tau}^2.
\label{ansatz}
\end{equation}
The value of the quark mass where the transition changes from 1st order to crossover, i.e.
the mass corresponding to the end-point, has been estimated in \cite{karschlat03} using
$N_{\tau}=4$ lattices to be $m^e=0.0007(4)$. This translates into the value of the 
pseudo-scalar mass
\begin{equation}
r_0 m_{ps}^e =0.16^{+3}_{-5}.
\end{equation}
It turns out that this large uncertainty in the value of $m_{ps}^e$ produces
an uncertainty in the extrapolated value of $T_c(m_{ps}^2)$ which is much smaller
than the statistical errors. The extrapolation according to Eq. (\ref{ansatz})
yields
\begin{equation}
r_0 T_c (m_{ps}^e)  = 0.429(8),~~\frac{T_c(m_{ps}^e)}{\sqrt{\sigma}}=0.391(9).
\end{equation}
For the fit to $r_0 T_c$ data we get $\chi^2/dof=0.7$, while for the fit to
$T_c(m_{ps}^e)/\sqrt{\sigma}$ data we have $\chi^2/dof=0.4$. The quark mass
dependence of the transition temperature is described by Eq. (\ref{ansatz})
only for $m>m^e$. For smaller quark masses the transition is first order and
$T_c$ depends linearly on the quark mass, i.e. we expect 
$T_c(m_{ps},N_{\tau})=T_c|_{cont}^{chiral} + A m_{ps}^2 + B/N_{\tau}^2$.
If we would insist on the linear dependence of the transition temperature
on the quark mass  in the entire mass range, the combined chiral and
continuum extrapolation would give
\begin{equation}
r_0 T_c  = 0.419(9),~~\frac{T_c}{\sqrt{\sigma}}=0.383(10).
\end{equation}
The $\chi^2/dof$ we have found for this fit is almost the same as for
the one above. The value of $T_c/\sqrt{\sigma}$ is slightly smaller than
the estimate of Ref. \cite{karsch01} based on $N_{\tau}=4$ lattice and larger
quark masses.
This is due to the continuum extrapolation performed in the present work.
Note, however, that the value $T_c(0)^{N_t=4}/\sqrt{\sigma}=0.417(9)$ is 
entirely consistent with Ref. \cite{karsch01}.
The value of $T_c$ could be compared with  the corresponding $2+1$ flavor value
$T_c r_0=0.444(6)[+12][-6]$ in the limit of vanishing $u$ and $d$ quark masses but fixed physical value of $m_s$  \cite{us}. Thus the flavor
dependence of $r_0 T_c$ is about or smaller than $5\%$. One should also note that
the difference between the transition temperature calculated on $N_{\tau}=4$ and
$N_{\tau}=6$ lattices is very similar to that found in 2+1 flavor case \cite{us}.
In Ref. \cite{karsch01} the transition temperature in the chiral limit has
also been estimated in units of the vector mass. Our estimate for
$T_c/m_V|_{m=0}$ is consistent with that result.

\section{Comparison of R and RHMC algorithm}

\begin{figure}
\includegraphics[width=7cm]{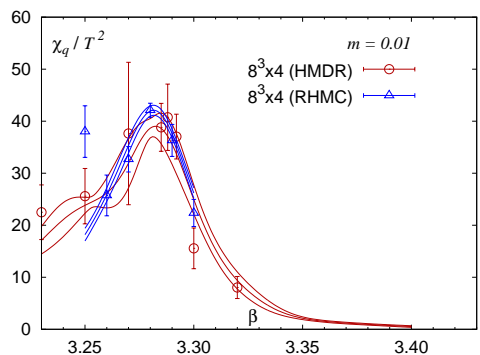}
\caption{The comparison of the disconnected part of chiral condensate susceptibility calculated with the R and RHMC algorithms
for p4fat3 with $m=0.01$.}
\label{fig:pbpsusc_alg_comp}
\end{figure}

We investigated the effect of the finite step-size errors of the R algorithm on the properties of
the finite temperature transition. We performed calculations on $8^3 \times 4$ lattices
using the p4fat3 action as well as the p4fat7 action and the later will be described in  
Appendix A in more detail.  
In the calculations with the p4fat3 action we used a quark mass of $m=0.01$, while
in case of p4fat7 action we used two quark masses $m=0.1$ and $0.035$.
In our calculations with the R algorithm the step-size of the molecular dynamics
evolution was set to be $dt=m/2.5$. Some additional calculations have been
done at twice smaller step-size $dt=m/5$. We have calculated the chiral condensate
and the Polyakov loop and determined the pseudo-critical couplings which are
summarized in Table \ref{tab:rcomp}.
In Fig. \ref{fig:pbpsusc_alg_comp} we compare the chiral 
condensate susceptibility  calculated using the R-algorithm and RHMC algorithm for 
the p4fat3 action.  We find that for the p4fat3 action the results obtained with R and RHMC algorithms
are identical within statistical errors.

\begin{figure}
\includegraphics[width=7cm]{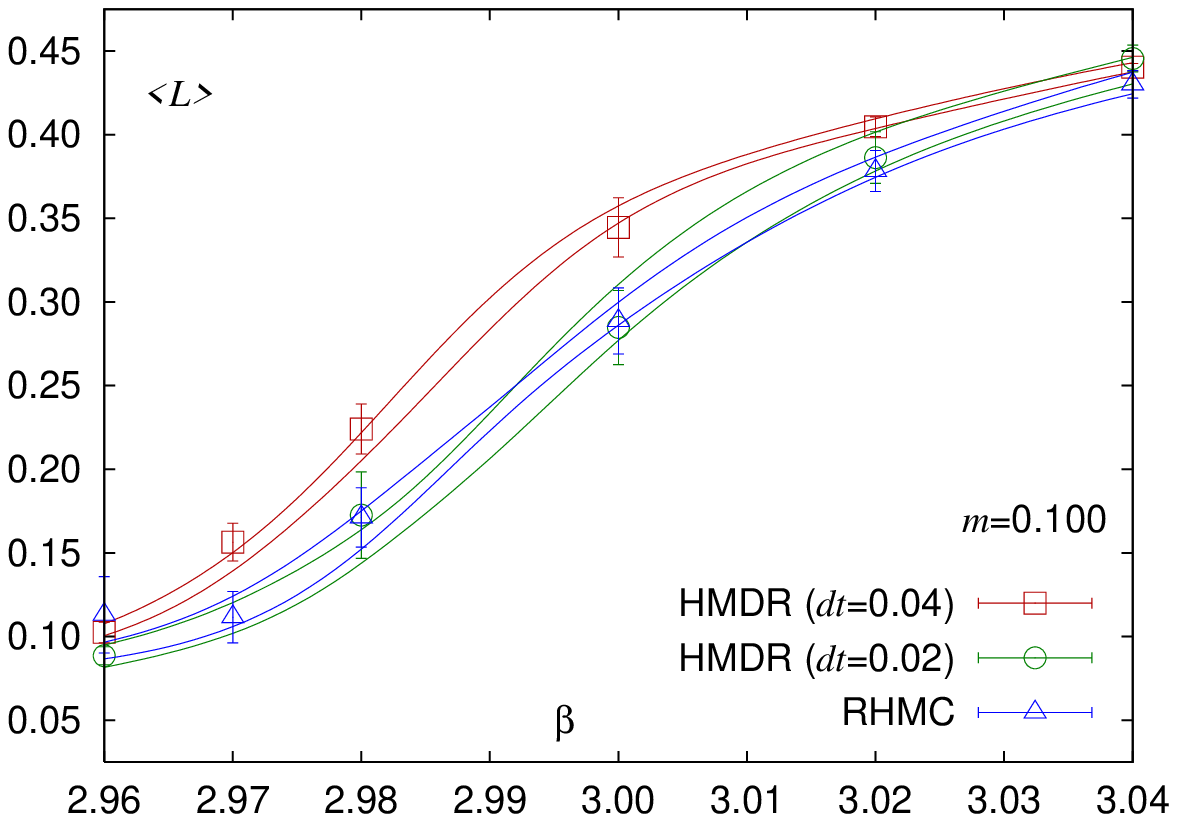}
\includegraphics[width=7cm]{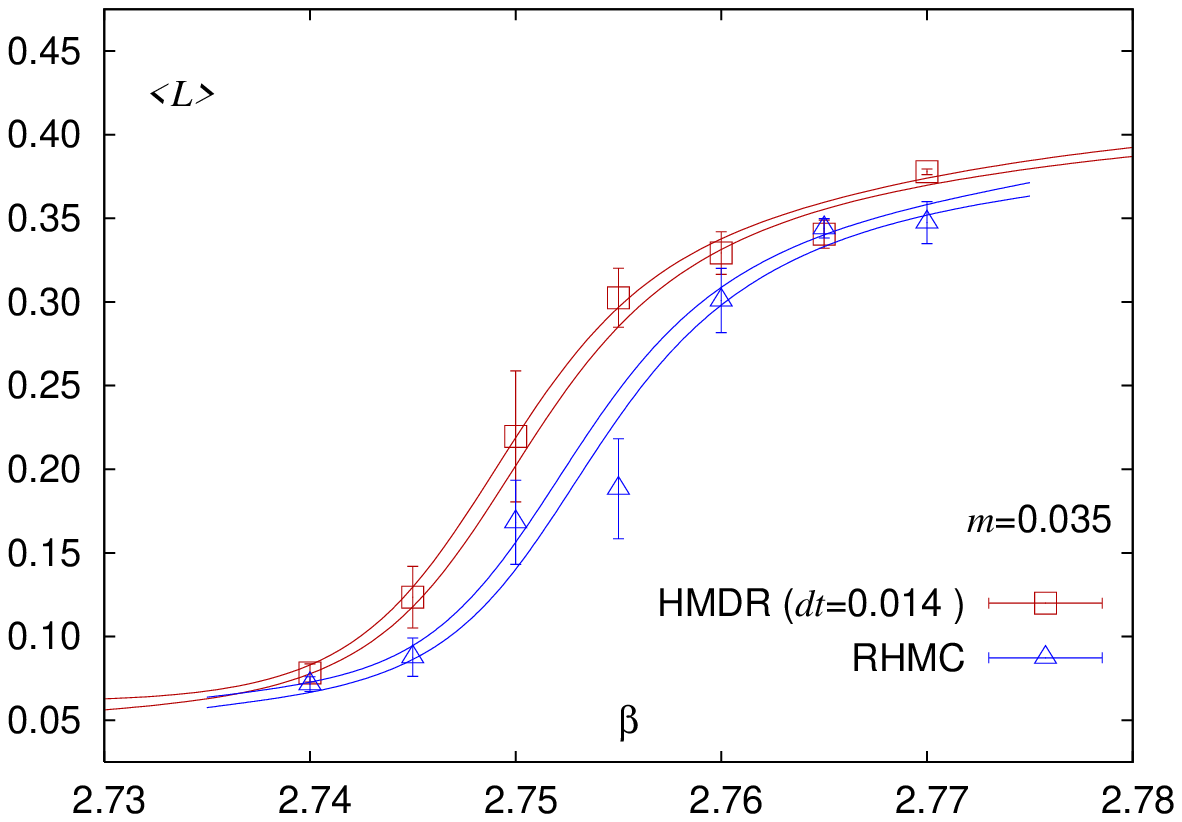}
\includegraphics[width=7cm]{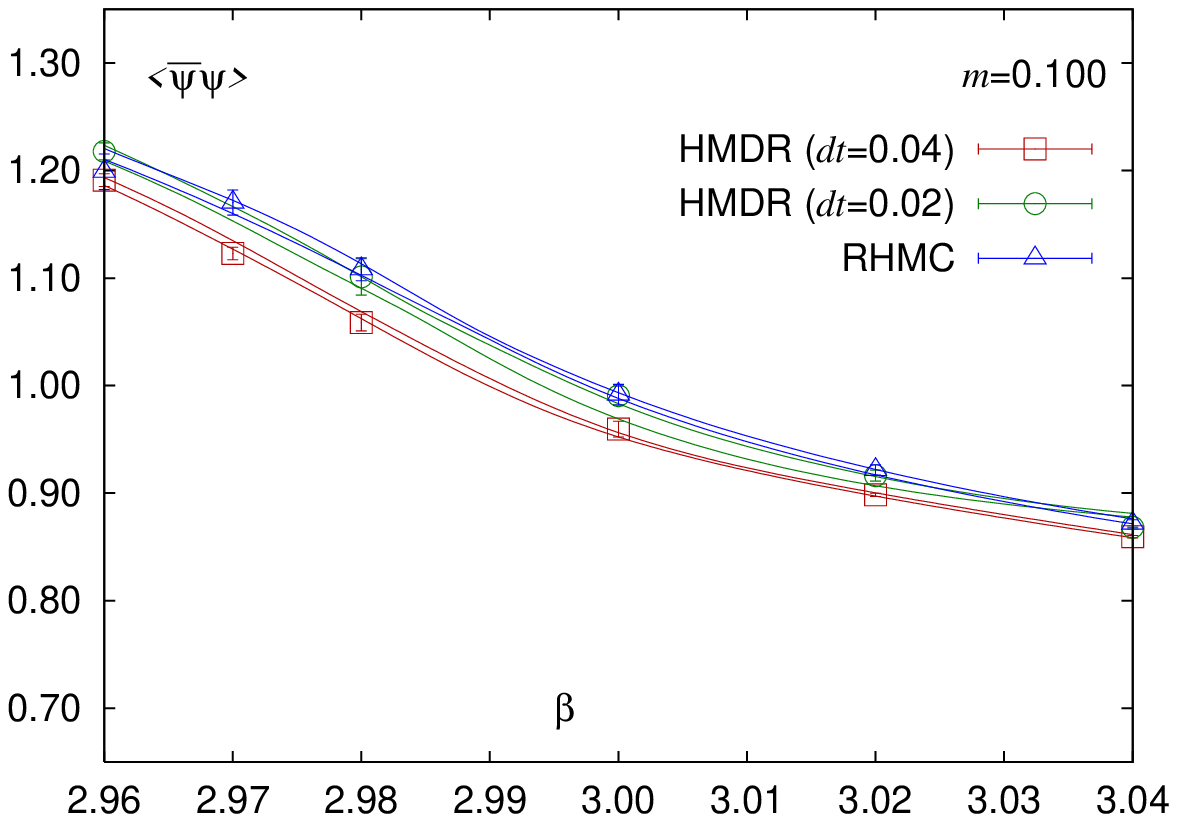}
\includegraphics[width=7cm]{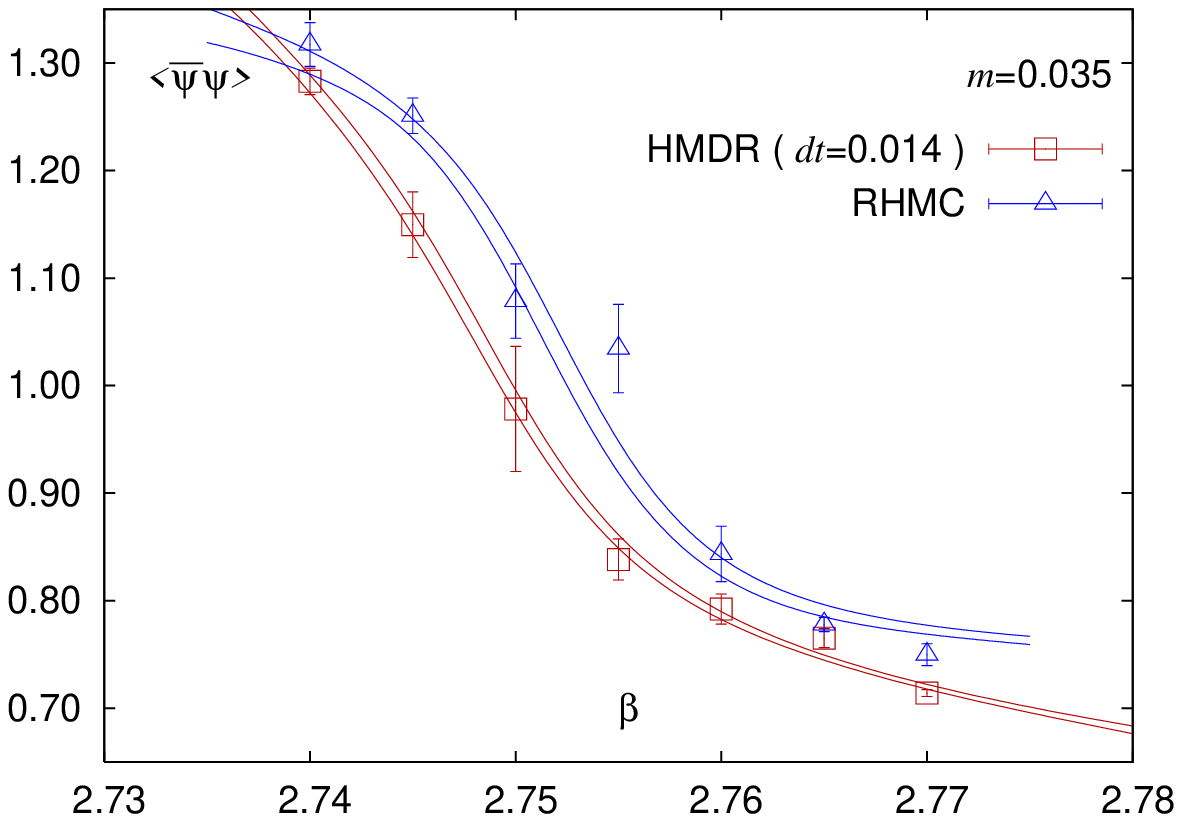}
\caption{The Polyakov loops (upper part) and the chiral condensate (lower part) calculated
with RHMC and R algorithms for the p4fat7 action with $m=0.1$ and $m=0.035$. }
\label{fig:fat7_alg_comp}
\end{figure}

The situation is different for
p4fat7. 
In Fig. \ref{fig:fat7_alg_comp} the expectation value of the Polyakov loop and the chiral
condensate calculated with the two algorithms are shown for two values of the quark mass.
Here we see significant differences
in the value of the chiral condensate and Polyakov loops calculated with the R algorithm and step-size 
$dt=m/2.5$ and the corresponding result obtained with RHMC algorithm. We see also
a small but statistically significant difference in the value of the pseudo-critical coupling calculated
with the two algorithms, c.f. Table \ref{tab:rcomp}. 
The difference becomes much less visible when the step-size is decreased
to $dt=m/5$. Figure \ref{fig:fat7_alg_comp} also suggests that the difference between the results of the
two algorithms becomes larger for the smaller quark mass.

For the p4fat7 action we also performed a zero temperature calculation on $16^3 \times 32$ lattices using the RHMC algorithm to determine the scale.
We have calculated meson masses as well the static quark potential. From the later we have extracted $r_0$. 
The results of these calculations  are also given in Table \ref{tab:rcomp}. Now we can estimate the transition temperature in units of $r_0$ for $m=0.035$
calculated with the two algorithms. For the R-algorithm we get $r_0 T_c=0.542(3)$ while for the RHMC algorithm  we have 
$r_0 T_c=0.552(3)$. Thus in the case of the p4fat7 action the R algorithm underestimates the transition temperature roughly by $2 \%$.
\begin{table}
\begin{tabular}{|c|c|c|c|c|c|c|}
\hline
Action  &  $m$    & Algorithm  & $\beta_{c, L}$   & $\beta_{c,q}$  &  $m_{ps}$  &   $r_0/a$   \\
\hline
p4fat3   &  0.010  & HMDR       & 3.2858(71)        &  3.2820(61)     &                   &                  \\
             &             & RHMC       & 3.2820(11)        &  3.2820(11)     &                   &                  \\
\hline
p4fat7   & 0.100  & HMDR      &  2.9850(25)       & 2.9753(53)       &                   &                  \\
             &             & RHMC      & 2.9939(33)        & 2.9831(20)       &                   &                  \\
\hline
p4fat7   & 0.035   & HMDR      &  2.7514(6)         & 2.7485(7)         &  0.7884(5)  &  2.1661(123) \\
             &             & RHMC      &  2.7540(6)         & 2.7515(7)         &  0.7897(7)  & 2.2063(108) \\
\hline
\end{tabular}
\caption{Comparison of the R and the RHMC algorithms for the pseudo-critical couplings and
the scale at the transition temperature.} 
\label{tab:rcomp}
\end{table}

\section{Conclusions}

In this paper we have studied the phase transition in 3 flavor QCD at finite
temperature using $N_{\tau}=4$ and $N_{\tau}=6$ lattices. For the quark mass
corresponding to the second order end-point we find the critical 
temperature to be $r_0 T_c=0.429(8)$. The transition temperature in the
chiral limit is about $2\%$ smaller than the above value. For a given 
pseudo-scalar meson mass the difference between the transition temperature in 
3 flavor and 2+1 flavor case is less than $5\%$. We also find that the cut-off
dependence of the transition temperature in 3 and 2+1 flavor QCD is very similar.
Furthermore, we find that finite step-size errors present in the R algorithm are
negligible, at least for the p4fat3 action at the quark masses studied.

\section*{Acknowledgments}
\label{ackn}
This work has been supported in part by contracts DE-AC02-98CH1-886 
and DE-FG02-92ER40699 with the U.S. Department of Energy, 
the Helmholtz Gesellschaft under grant
VI-VH-041 and the Deutsche Forschungsgemeinschaft under grant GRK 881.  
The work of C.S. has been supported through LDRD funds from Brookhaven
National Laboratory.
The majority of the calculations reported here were carried out using
the QCDOC supercomputers of the RIKEN-BNL Research Center and the U.S.
DOE.  In addition some of the work was done using the APE1000 supercomputer
at Bielefeld University

\section*{Appendix A} 
In this appendix we are going to discuss the
properties of the finite temperature transition in the case of the p4fat7
action and compare it to the p4fat3 case.  In general the gauge transporter
in the 1-link term of the p4 action can be replaced by combination of
the link variable and different staples, called the fat link, without
changing the naive continuum limit. This is true provided the coefficient
of different terms in the fat link satisfy appropriate normalization
conditions.  For example in the case of the fat link with the three link
staple only, this condition reads $c_1 +6 c_3=1$.  Introducing five
and seven link staples  in addition to the three link staples give
the so-called fat7 link \cite{kostas}.  In this case the normalization
condition reads: $c_1+6c_3+24 c_5+48 c_7=1$. It is possible to
eliminate the leading order coupling to the high momentum gluons with
momenta $(0,\pi,0,0)$, $(0,\pi,\pi,0)$ and $(0,\pi,\pi,\pi)$, i.e. to
suppress the flavor changing interaction at order $g^2 a^2$ if the
coefficients are chosen as \cite{kostas}
\begin{equation}
\frac{c_3}{c_1}=\frac{1}{2},~\frac{c_5}{c_1}=\frac{1}{8},~\frac{c_7}{c_1}=\frac{1}{48}.
\end{equation}
This gives then the value $c_1=-1/8$ for the coefficient of the 1- link term to get
the naive continuum limit.

\begin{figure}
\includegraphics[width=7cm]{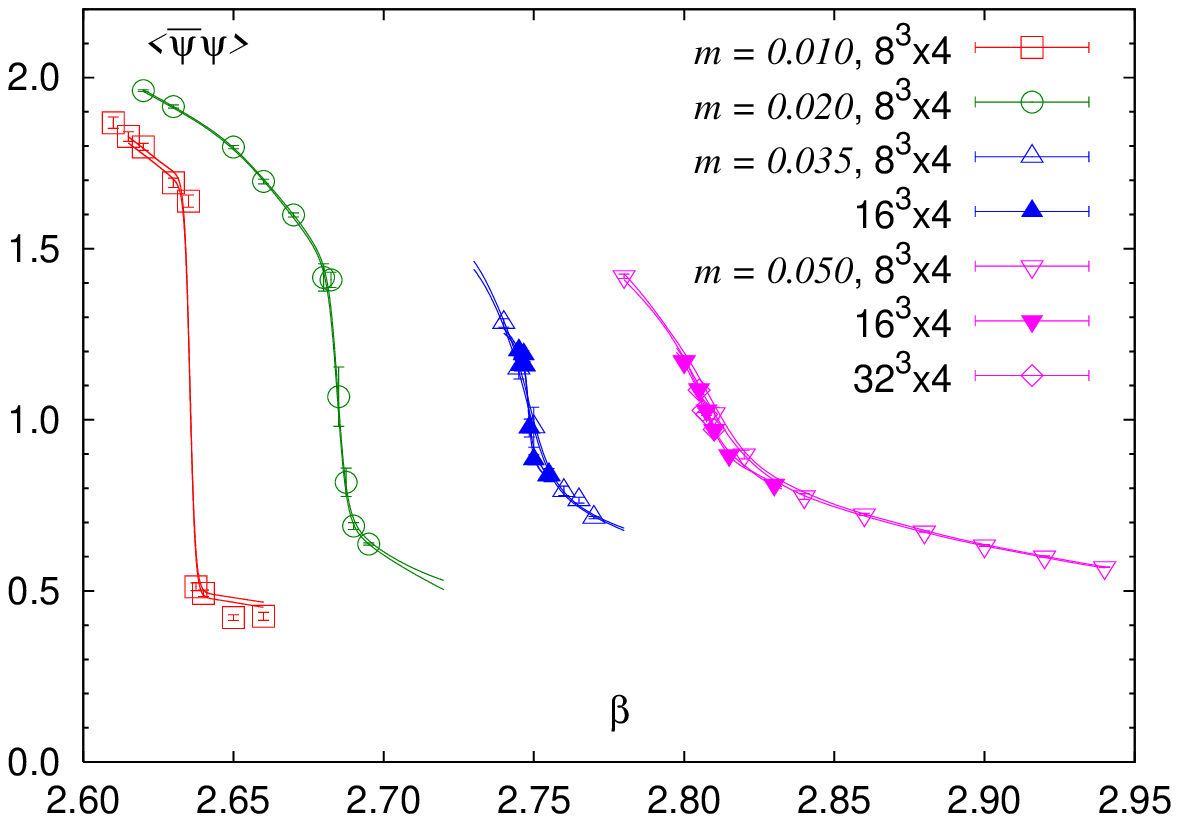}
\includegraphics[width=7cm]{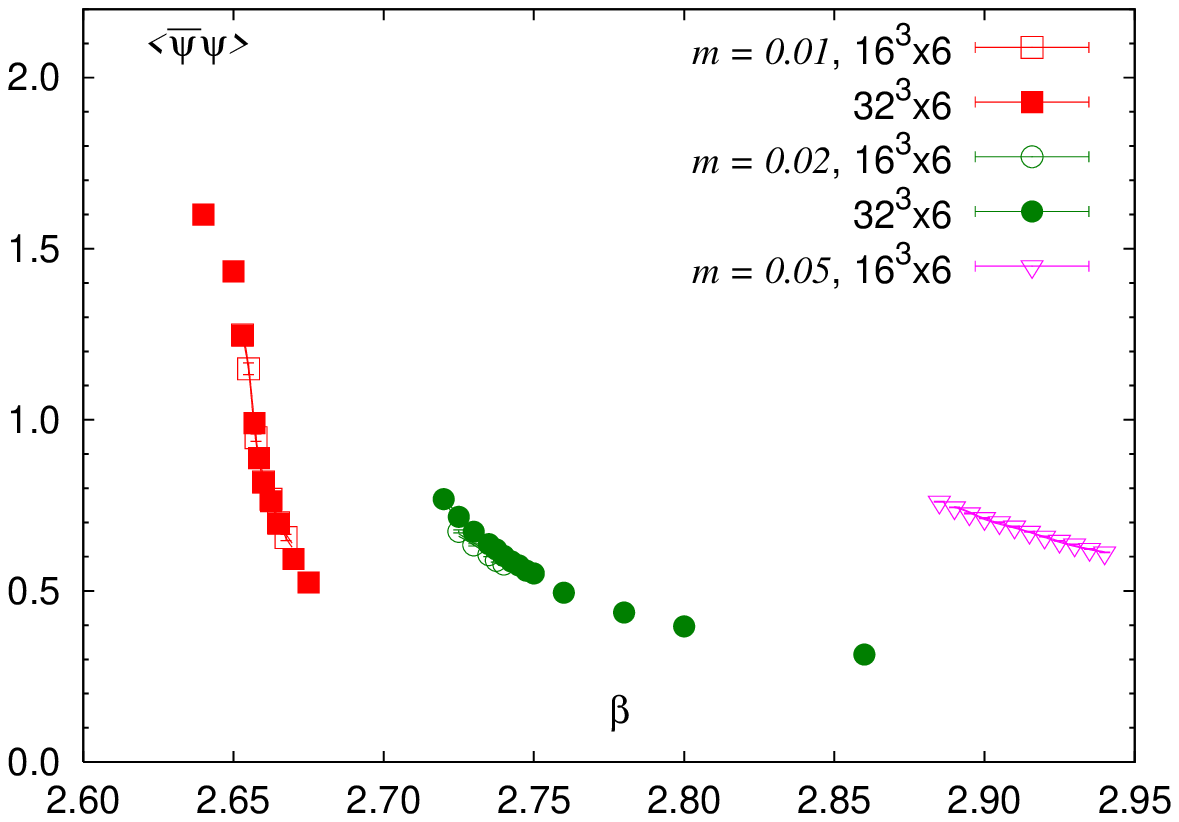}
\caption{The chiral condensate calculated on $N_{\tau}=4$ (left) and
$N_{\tau}=6$ (right) lattices for p4fat7 action.}
\label{fig:pbpfta7}
\end{figure}

\begin{figure}
\includegraphics[width=7cm]{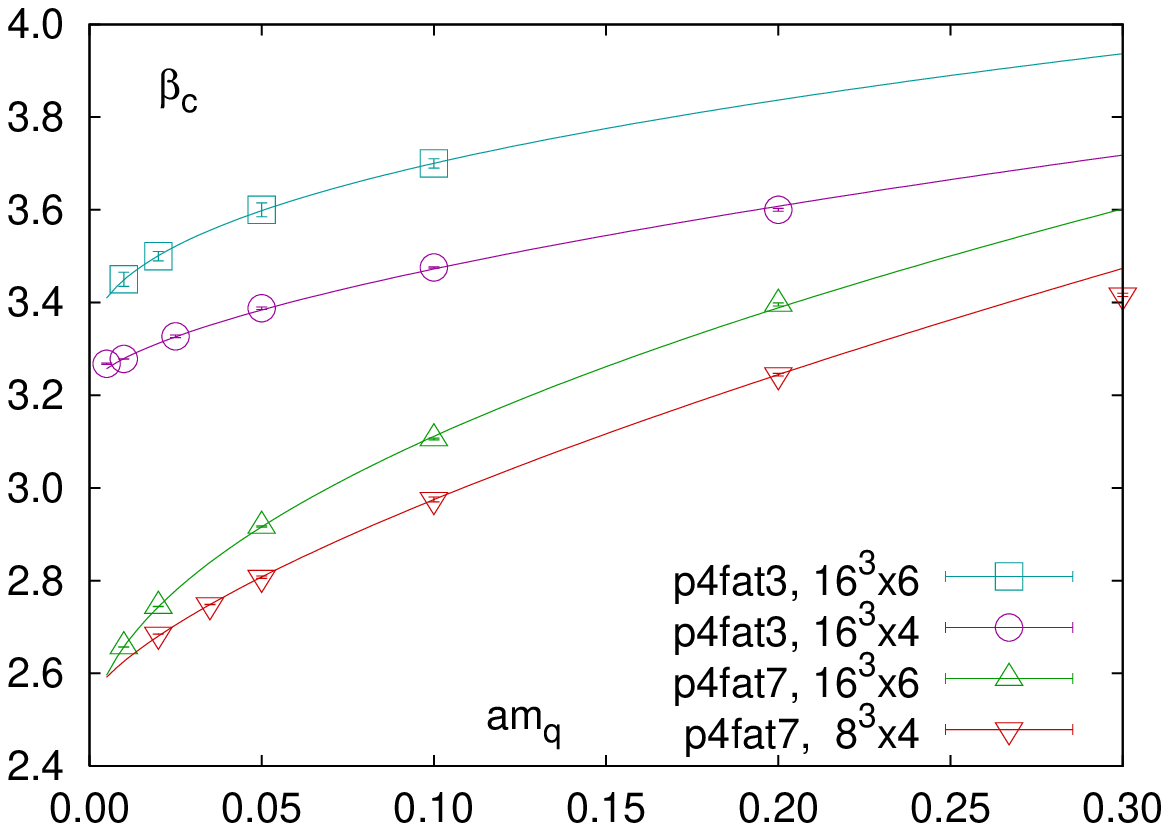}
\includegraphics[width=7cm]{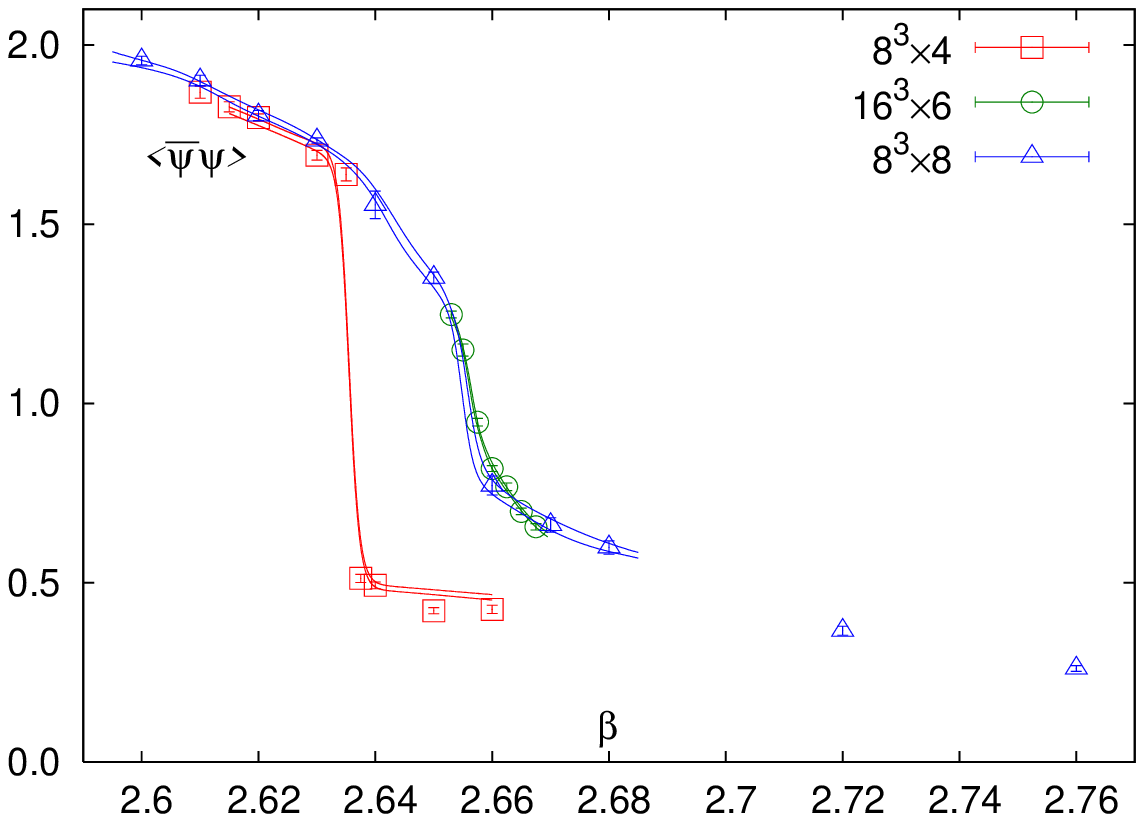}
\caption{The pseudo-critical couplings for the p4fat3 and p4fat7 actions (left) and the
chiral condensate calculated with the p4fat7 action on $8^3 \times 4$, $16^3 \times 6$ and
$8^4$ lattices at $m=0.01$ (right). 
}
\label{fig:betac} 
\end{figure}

In Fig. \ref{fig:pbpfta7} we show the chiral condensate calculated on
the $N_{\tau}=4$ and $N_{\tau}=6$ lattices.  We can see that for small
quark masses the transition becomes strongly first order. The value of
the chiral condensate in the low temperature phase is also much larger
than for the p4fat3 action discussed in the main text. For the
smallest quark mass the discontinuity in the chiral condensate is
about the same for $N_{\tau}=4$ and $N_{\tau}=6$ lattice, although we would
expect it to decrease by roughly a factor of $(6/4)^3 \simeq 3$ when
going from $N_{\tau}=4$ to $N_{\tau}=6$.  This could mean that we are dealing
with a bulk transition.  In Fig. \ref{fig:betac} we also compare the
pseudo-critical couplings for the p4fat3 and p4fat7 actions. We see
that for large quark masses the $N_{\tau}$ dependence of the
pseudo-critical coupling is similar, though their values are
significantly different.  For small quark masses the pseudo-critical
couplings calculated for $N_{\tau}=4$ and $6$ come very close
together, again suggesting that the transition may be a bulk
transition. We did calculations also on $8^4$ lattice at $m=0.01$.
In Fig. \ref{fig:betac} we compare the chiral condensate calculated on $8^3 \times 4$, $16^3 \times 6$ and
$8^4$ lattices. We see a sharp drop in the value of the chiral condensate, which occurs at the same $\beta_c$ for
$N_{\tau}=6$ and $8$. This again indicates a bulk transition.

\begin{figure}
  \includegraphics[width=7cm]{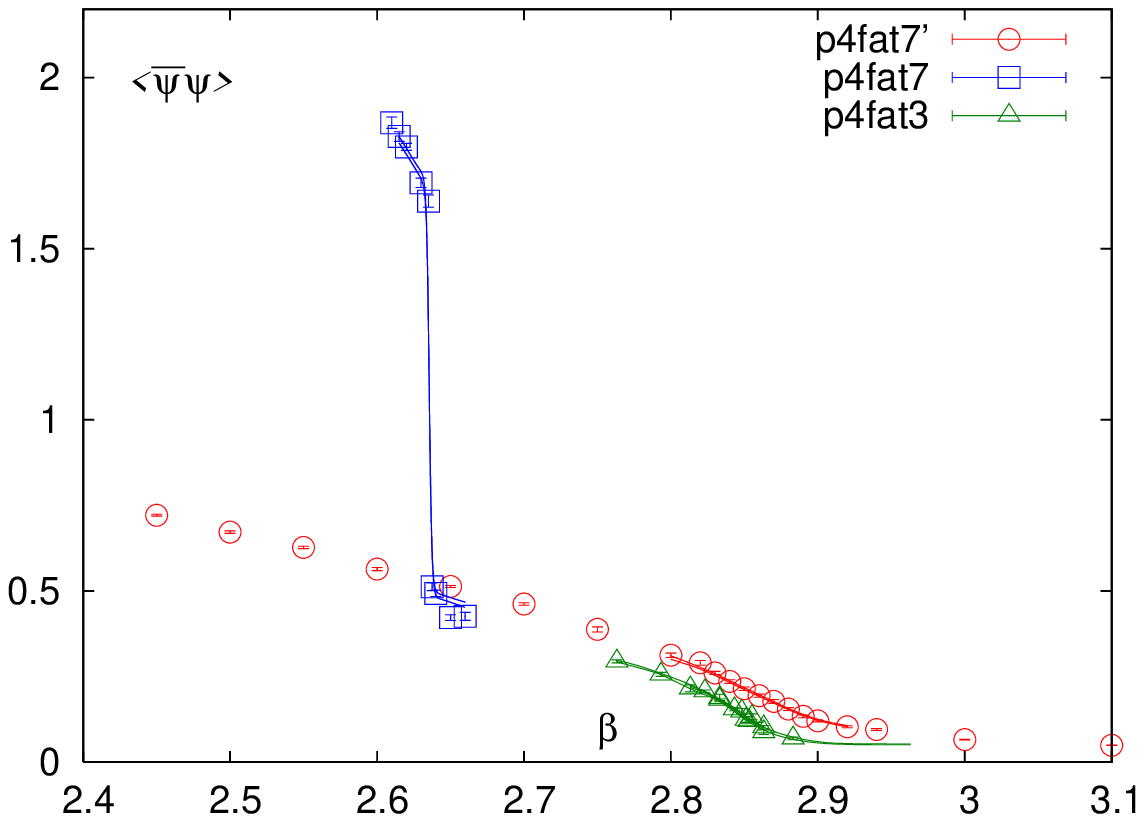}
  \includegraphics[width=7cm]{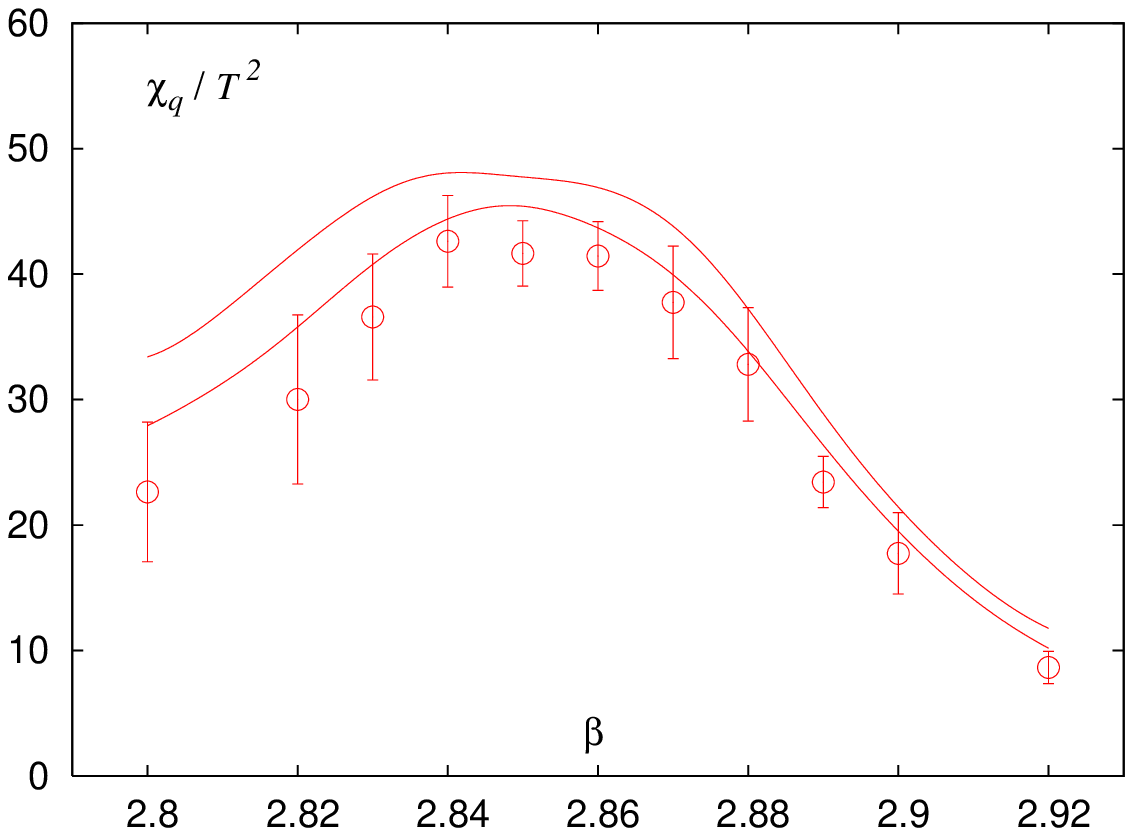}
  \caption{The chiral condensate (left) and its susceptibility (right)
    for the p4fat7' action calculated on $8^3 \times 4$ lattice at $m=0.01$. We also show the chiral condensate
calculated with p4fat7 and p4fat3 actions at the same quark mass. The data points for p4fat3 action have been shifted horizontally by $\Delta\beta=-0.437$ 
for better visibility. The band in the right figure corresponds to Ferrenberg Swendsen re-weighting.}
\label{fig:pbpfat7p}
\end{figure}

One may wonder which feature of the p4fat7 action is responsible for
the bulk transition.  The main difference of the p4fat7 action
compared to p4fat3 action as well as to other fat link action (e.g.
ASQTAD) is the negative sign of the one link term. 
Close to the continuum limit the normalization condition
$c_1 + 6 c_3 + 24 c_5 + 48 c_7 = 1$ should insure that the
combination of 1-, 3-, 5- and 7-link terms will describe a
conventionally normalized, positive Dirac kinetic energy.
However, at stronger coupling where the gauge field are
more disordered, the staples with many links are expected
to give a significantly smaller contribution and the 1-link
term may dominate, resulting in an effective kinetic energy
term with a possibly negative sign.  While this would
simply correspond to non-standard sign and normalization
conventions for the Dirac kinetic energy, it raises the
possibility that this effective kinetic energy term will
change sign as one passes from strong to weak coupling.
Such a sign change could induce a bulk transition.  In addition,
the change in magnitude of the coefficient in the effective
kinetic energy (small for strong coupling and large for
weak coupling) would appear reversed in the chiral condensate
(large for strong coupling and small for weak coupling)
consistent with the observed behavior.
To verify this we did calculations with p4fat7 action but with
different coefficients which we call the p4fat7' action.  The
coefficients were chosen to be
\begin{equation}
  c_1=\frac{3}{4} \cdot  \frac{1}{8},~\frac{c_3}{c_1}=\frac{1}{2},~\frac{c_5}{c_1}=\frac{1}{8},~\frac{c_7}{c_1}=\frac{1}{48}.
\end{equation}         
For this action we found no evidence for a strong first order
transition but only a crossover. This can be seen for example in the
behavior of the chiral condensate shown in Fig. \ref{fig:pbpfat7p}.
Both the value of the chiral condensate and the location of the
transition point is very similar to that of the p4fat3 action.

\begin{figure}
\includegraphics[width=5.5cm]{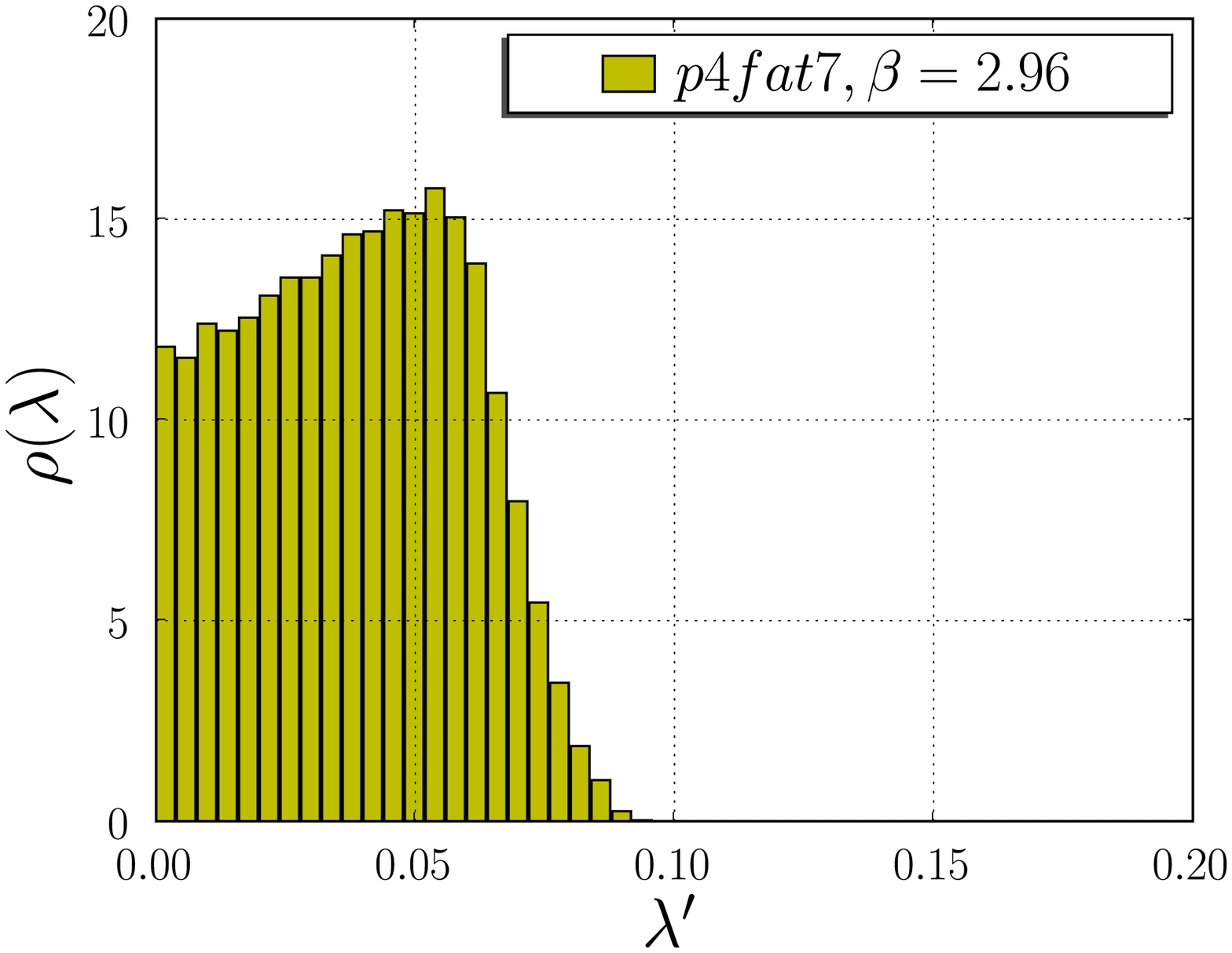}
\includegraphics[width=5.5cm]{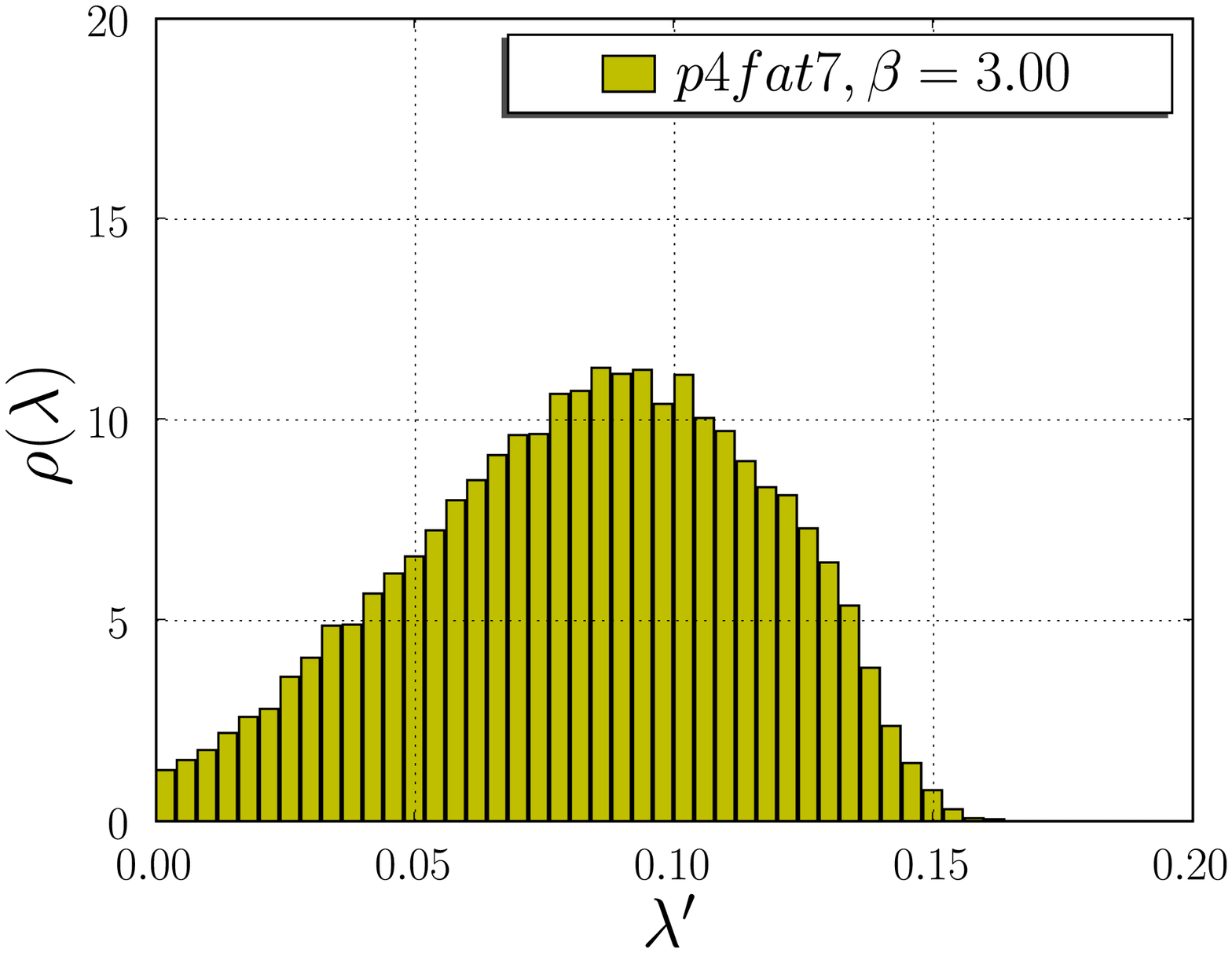}
\includegraphics[width=5.5cm]{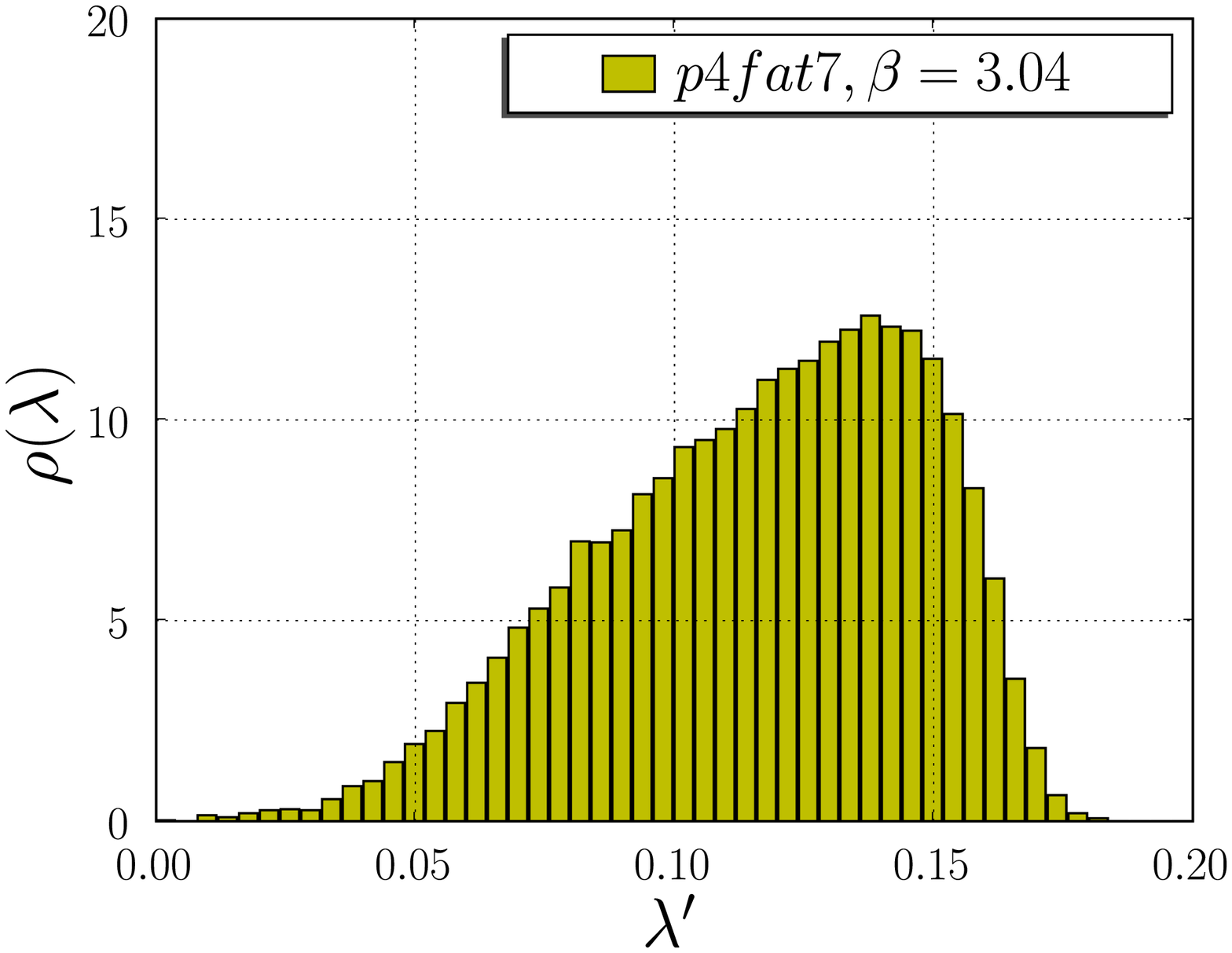}
\caption{The eigenvalue distributions of the p4fat7 Dirac operator for
  $m=0.1$ calculated on $8^3 \times 4$ lattice in the confined phase
  ($\beta=2.96$), at the transition ($\beta=3.00$) and in the
  deconfined phase ($\beta=3.04$).  }
\label{fig:eig_m01}
\end{figure}

\begin{figure}  
\includegraphics[width=5.5cm]{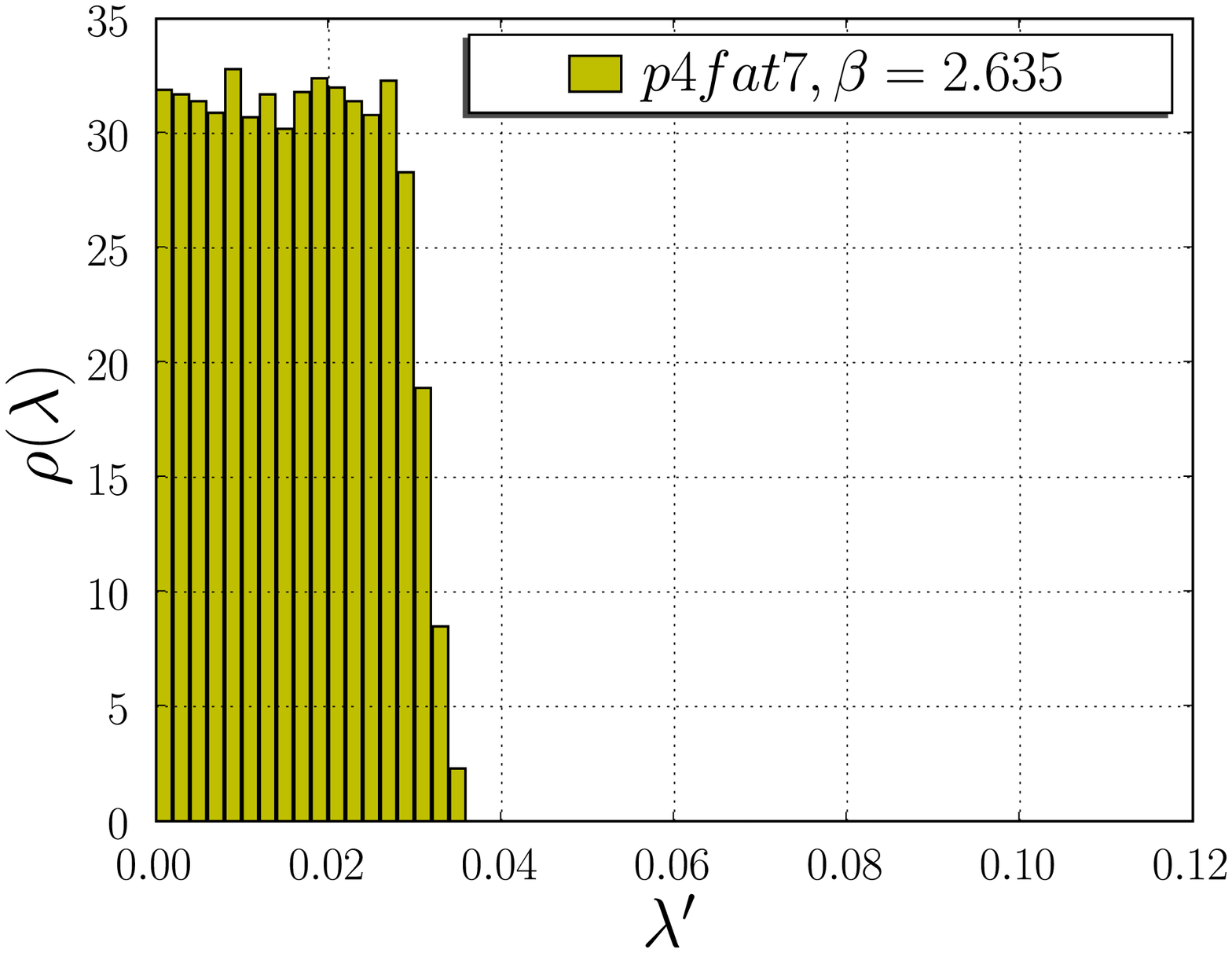}              
\includegraphics[width=5.5cm]{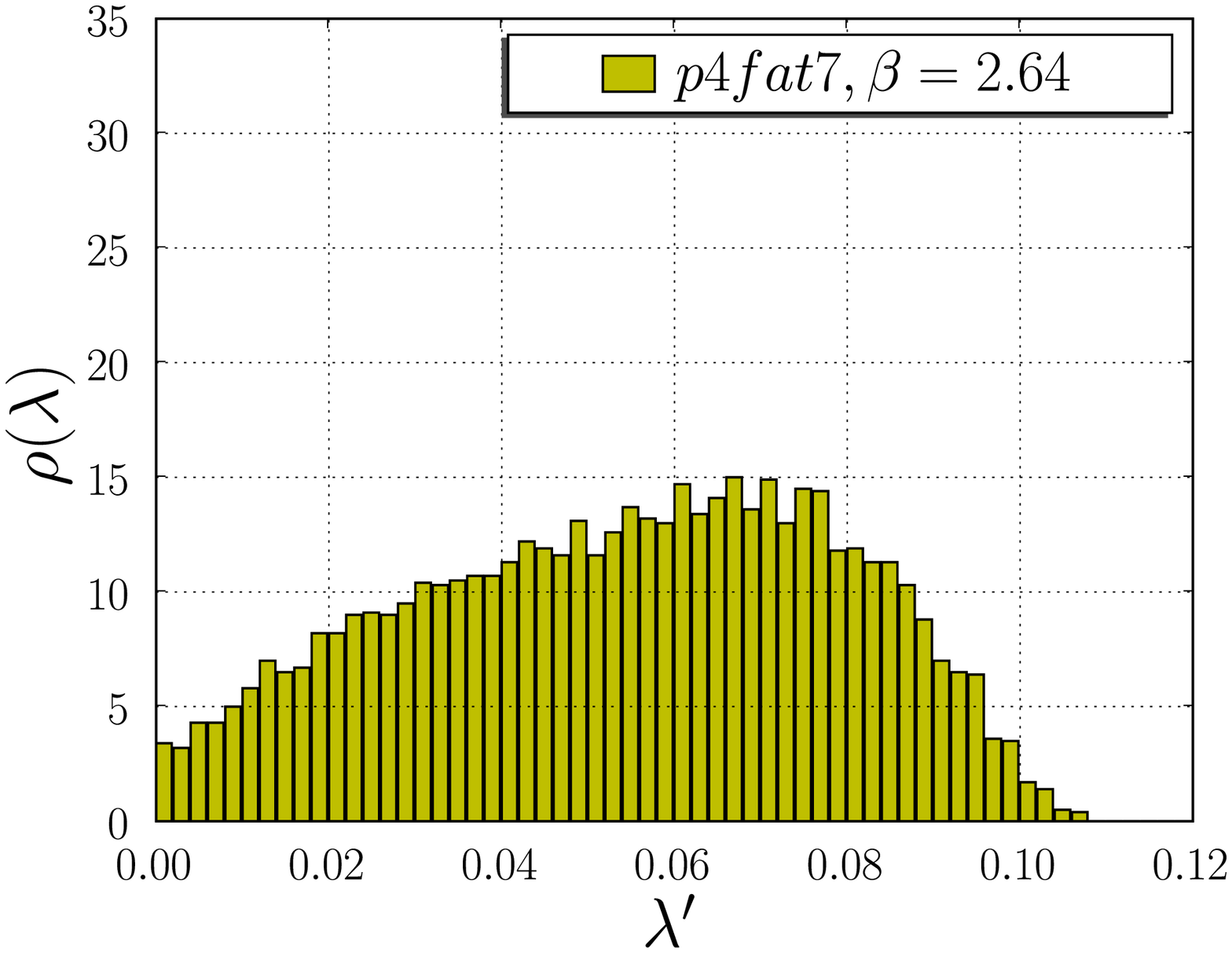}
\includegraphics[width=5.5cm]{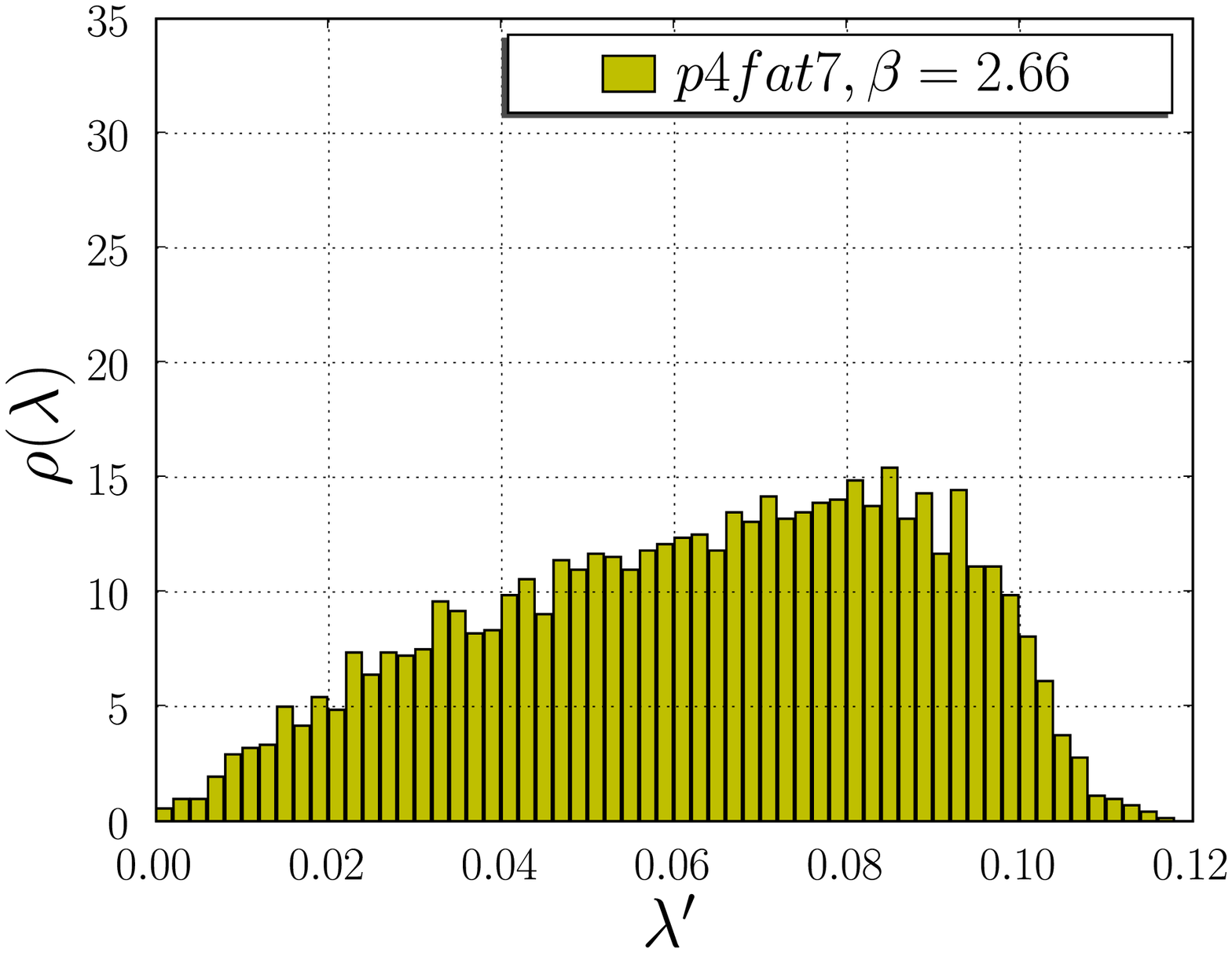}
\caption{The eigenvalue distributions of the p4fat7 Dirac operator for
  $m=0.01$ calculated on $8^3 \times 4$ lattice in the confined phase
  ($\beta=2.635$), at the transition ($\beta=2.64$) and in the
  deconfined phase ($\beta=2.66$).  }
\label{fig:eig_m001}
\end{figure}

We also calculated the eigenvalues $\lambda=\imath \lambda' +2 m$ of the p4fat7
Dirac operator. The normalized distribution of the lowest 50
eigenvalues $\lambda'$ is shown in Figs.  \ref{fig:eig_m01} and
\ref{fig:eig_m001} for the quark masses $m=0.1$ and $m=0.01$,
respectively. We have used 100 configurations for $m=0.01$, and 200
configurations for $m=0.1$.  
Given the above definition of $\lambda'$ the breaking of the chiral
symmetry manifests itself in a non-zero density at $\lambda' \simeq 0$.
In Fig. \ref{fig:eig_m01} we show the eigenvalue distribution for the
larger quark mass $m=0.1$.  It shows the expected features: large
density of eigenvalues near $\lambda \simeq 0$ in the low temperature
phase ($\beta=2.96$), significant drop of eigenvalue density around
zero at the transition ($\beta=3.0$) and zero density of eigenvalues
at the origin for the deconfined phase ($\beta=3.04$).  The situation
is different for the smallest quark mass $m=0.01$, where we see
non-zero density of eigenvalues at $\lambda \simeq 0$ even in the
deconfined phase ($\beta=2.66$). The large decrease in the density of
eigenvalues at the origin when going from the confined phase
($\beta=2.635$) to the deconfined explains the large drop in the value of
the chiral condensate.

\section*{Appendix B}
In this appendix we discuss the calculations of the largest and the
smallest eigenvalue of the staggered Dirac operator in the free field
limit. Let us start our discussion with case of the standard staggered
fermions.  The free-field staggered Dirac operator acting on a
single-component fermion field is given by
\begin{eqnarray}
  \mathcal{M} \chi(x) = \sum_{\mu}\eta_{\mu}(x)\left(\chi(x+\mu) - 
    \chi(x-\mu) \right) + 2m\chi(x),
\end{eqnarray}
where $\eta_{\mu}(x) = (-1)^{x_0 + x_1 + \ldots x_{\mu-1}}$ are the
normal staggered phases. Consider $\mathcal{M}$ acting on a momentum
eigenstate
\begin{eqnarray}
  \chi(x) & = & \chi(p) \, \mathrm{e}^{\imath p \cdot x} \label{chifield}, \\
  \mathcal{M} \chi(x) & = & \sum_{\mu}\left[\chi(p) \,
    \mathrm{e}^{\imath p \cdot x}\,
    (\mathrm{e}^{\imath p_{\mu}} - \mathrm{e}^{-\imath p_{\mu}}) \, 
    \mathrm{e}^{\pi_{\mu} \cdot x} \right]
  + 2m \,\chi(p) \,\mathrm{e}^{\imath p \cdot x},
\end{eqnarray}
where $\pi_{0} = (0,0,0,0), \pi_{1} = (\pi, 0, 0, 0), \pi_{2} = (\pi,
\pi, 0, 0), \ldots$ Thus, the staggered Dirac operator has
non-diagonal terms that couple together states at different corners of
the Brillouin zone.

In momentum space, we can write the fermion matrix as
\begin{eqnarray}
  \mathcal{M}_{p',p} & = & \sum_{\mu} 2\imath\sin(p_{\mu}) \,
  \delta_{p',p+\pi_{\mu}} + 2m \,\delta_{p',p} \\
  \mathcal{M^{\dagger}}_{p'',p'} & = & \sum_{\mu} 
  -2\imath\sin(p''_{\mu})\, \delta_{p''+\pi_{\mu}, p'} +
  2m \,\delta_{p'',p'}.
\end{eqnarray}
Then, we have for $\mathcal{M}^{\dagger}\mathcal{M}$
\begin{eqnarray}
  \mathcal{M^{\dagger}}_{p'',p'}\mathcal{M}_{p',p}  & = 
  & 4 \sum_{\mu}\sum_{\nu}\sin(p_{\mu}) \sin(p''_{\nu}) \,
  \delta_{p''+\pi_{\nu}, p+\pi_{\mu}} + 4m^2 \delta_{p,p''} \\
  & = & 4 \sum_{\mu}\sum_{\nu}\sin(p_{\mu})
  \sin(p_{\nu}+(\pi_{\mu})_{\nu} - (\pi_{\nu})_{\nu}) \, \delta_{p''+
    \pi_{\nu}, p+\pi_{\mu}} +  4m^2 \delta_{p,p''} \\
  & = & 4\sum_{\mu}\sum_{\nu<\mu}\sin(p_{\mu})
  \sin(p_{\nu}+\pi) \,\delta_{p'',p+\pi_{\mu}-\pi_{\nu}} +  \nonumber\\
  & & 4\sum_{\mu}\sum_{\nu>\mu}\sin(p_{\mu})\sin(p_{\nu}) \,\delta_{p'',p+
    \pi_{\mu}-\pi_{\nu}} +  \nonumber \\
  &  & 4\sum_{\mu} \sin^2(p_{\mu}) \, \delta_{p,p''} +
  4m^2  \delta_{p,p''}. \label{eq:intmunu}
\end{eqnarray}
Noticing that $p+\pi_{\mu}-\pi_{\nu} = p + \pi_{\nu}-\pi_{\mu}$ (mod
2$\pi$), and interchanging $\mu$ and $\nu$ labels in the second piece
of Eq.~(\ref{eq:intmunu}), we see that all the off-diagonal pieces of
$\mathcal{M}^{\dagger}\mathcal{M}$ cancel, and we are left only with
the diagonal piece,
\begin{eqnarray}
  (\mathcal{M}^{\dagger}\mathcal{M})_{p'',p} = 
  4\sum_{\mu}\sin^2(p_{\mu}) \, \delta_{p,p''} +
  4m^2 \delta_{p,p''}.
\end{eqnarray}
As expected, we have a hard lower bound on the eigenvalue spectrum of
$\lambda^2_{min}=4m^2$.  The upper bound $\lambda^2_{max} = 16 + 4m^2$ is
realized when $p =
(\frac{\pi}{2},\frac{\pi}{2},\frac{\pi}{2},\frac{\pi}{2})$.
 
The situation for free-field p4 fermions is similar, but a little bit
more complicated.  Here, the p4 Dirac operator is given by
\begin{eqnarray}
  \mathcal{M} \chi(x) & = & 2m\chi(x) + c_{1,0}\sum_{\mu}\eta_{\mu}(x) 
  \left( \chi(x+\mu) - \chi(x-\mu) \right)+ {}  \\
  & & + c_{1,2}\sum_{\mu}\sum_{\nu \neq \mu}\eta_{\mu}(x)
  \left(\chi(x+\mu+2\nu)+\chi(x+\mu-2\nu)-
    \chi(x-\mu+2\nu)-\chi(x-\mu-2\nu)\right),\nonumber
\end{eqnarray}
where $c_{1,0}=\frac{3}{4}$ and $c_{1,2}=\frac{1}{24}$. Again, we can
examine how $\mathcal{M}$ acts on the momentum eigenstate in
Eq.~(\ref{chifield}).  As before, we see that we have off-diagonal
pieces that come about as a direct result of the presence of staggered
phases.
\begin{eqnarray}
  \mathcal{M} \chi(x) & = & \sum_{\mu}\chi(p)\,
  \mathrm{e}^{\imath (p+\pi_{\mu}) \cdot x} \,\imath \, h_{\mu}(p) + 
  2m \,\chi(p) \, \mathrm{e}^{\imath p \cdot x}, \\
  h_{\mu}(p) & = & 2c_{1,0}\sin(p_{\mu}) + 
  4c_{1,2}\sum_{\nu \neq \mu}\sin(p_{\mu})\cos(2p_{\nu}).
\end{eqnarray}
Or, in matrix form,
\begin{eqnarray}
  \mathcal{M}_{p',p} & = & \sum_{\mu} \imath h_{\mu}(p) \,
  \delta_{p',p+\pi_{\mu}} + 2m \, \delta_{p',p}.
\end{eqnarray}
Calculating $\mathcal{M}^{\dagger}\mathcal{M}$, we see that the
off-diagonal pieces are eliminated in the same way as in the naive
staggered case. Thus, we are left with only diagonal terms,
\begin{eqnarray}
  (\mathcal{M}^{\dagger}\mathcal{M})_{p'',p} =  \sum_{\mu}h_{\mu}^2(p) \,
  \delta_{p,p''} + 4m^2 \delta_{p,p''}.
\end{eqnarray}
We see that $\lambda^2_{min} = 4m^2$.  Finding $\lambda^2_{max}$ requires
us to maximize the function $\sum h^2_{\mu}(p)$.  Doing this, we find
the maximum when two of the components of $p$ are equal to $\pi/2$ and
the other two components are equal to $0$.  For example, $p_{max} =
(\pi/2,0,\pi/2,0)$.  This yields $\lambda^2_{max} = 50/9+4m^2$ for the
p4 action.  A similar calculation for the Naik action yields
$\lambda^2_{max} = 196/9+4m^2$.

For completeness, we also quote the eigenvectors of the free p4 Dirac
operator. Using a slightly different method we find,
\begin{equation}
\label{eq:ansatzup4}
\chi(x) \equiv \chi_\rho(X) = \left[ \imath \sum_\mu
  \Gamma_{\rho\rho'}^\mu(p) \, \sin p_\mu \left(
  2 c_{1,0}+ 4 c_{1,2} \sum_{\nu \ne \mu}\cos 2 p_\nu 
  \right) + (\lambda -2 m) \, \delta_{\rho\rho'} \right] u^0_{\rho'} \; 
\mathrm{e}^{\imath 2 p \cdot X},
\end{equation}
where we implicitly sum over $\rho'$. Here we have used hypercube
coordinates $X$ and $\rho$, where $X$ labels the hypercube and $\rho$
is the offset within the hypercube, $x = 2 X + \rho$ with
$\rho_\mu=0,1$.  Furthermore,
$\Gamma_{\rho\rho'}^\mu(p)$ is defined by
$\Gamma_{\rho\rho'}^\mu(p) :=\eta_\mu(\rho)\; [
\delta_{(\rho+\hat{\mu}),\rho'} + \delta_{(\rho-\hat{\mu}),\rho'}]
\mathrm{e}^{\imath p \: (\rho-\rho')}$ and $u_{\rho}^0$ is a constant
vector depending only on $\rho$.  Finally,  the eigenvalue of
the free p4 Dirac operator is
\begin{equation}
  \label{eq:eigp4full}
  \lambda = 2 m \pm \imath \sqrt{\sum_{\mu} \sin^2 p_\mu
  \left( 2 c_{1,0} + 4 c_{1,2}\sum_{\nu \ne \mu}
    \cos 2 p_\nu \right)^2 }.
\end{equation}

\end{document}